\documentclass[a4paper,11pt]{article}
\pdfoutput=1 

\usepackage{jcappub} 

\usepackage[T1]{fontenc} 

\graphicspath{{Plots/}}

\title{Do assumptions about the central density of subhaloes affect dark matter annihilation and lensing calculations?}

\author[a]{Nicole E. Drakos,\note{Corresponding author.}}
\author[b,c]{James E. Taylor,}
\author[d]{and Andrew J. Benson}

\affiliation[a]{Department of Physics and Astronomy, University of Hawaii, \\ Hilo, 200 W Kawili St, Hilo, HI 96720, USA}
\affiliation[b]{Department of Physics and Astronomy, University of Waterloo, \\ 200 University Avenue West, Waterloo, ON N2L\,3G1, Canada}
\affiliation[c]{Waterloo Centre for Astrophysics, University of Waterloo, \\ 200 University Avenue West, Waterloo, ON N2L\,3G1, Canada}
\affiliation[d]{Carnegie Observatories,\\ 813 Santa Barbara Street, Pasadena, CA 91101, USA}

\emailAdd{ndrakos@hawaii.edu}
\emailAdd{taylor@uwaterloo.ca}
\emailAdd{abenson@carnegiescience.edu}

\abstract{Subhalo models play a critical role in dark matter annihilation predictions and galaxy-galaxy lensing studies; however the internal structure of subhaloes remains highly uncertain. In particular, a growing body of evidence suggests that the central density of cuspy dark matter subhaloes is conserved in minor mergers, whereas empirical models of subhalo evolution---calibrated using limited-resolution simulations---often assume a drop in the central density. To assess the impact of these assumptions, we systematically explore how a wide range of initial mass profiles and tidal evolution prescriptions influence annihilation and lensing calculations, including the physically motivated Energy Truncation model, which explicitly preserves the central density of subhaloes. We find that annihilation calculations are very sensitive to the assumed inner density profile, and different models can  produce more than an order of magnitude difference in the annihilation rate of individual subhaloes,
 and a factor of $\sim5$ in the total annihilation rate expected in the Milky Way. Since the innermost parts of haloes will always be difficult to resolve in simulations,  we conclude that developing a theoretical understanding of subhalo evolution is crucial to be able to make accurate predictions of the dark matter annihilation signal. On the other hand,  while the shear and convergence profiles used in galaxy-galaxy lensing are sensitive to the initial profile assumed (e.g., NFW versus Einasto),  they are otherwise well-approximated by a simple stripping model in which the original profile is sharply truncated at a tidal radius.}

\keywords{dark matter theory,  semi-analytic modeling, gamma ray theory}

\begin{document}
\maketitle
\flushbottom
	
\section{Introduction} \label{sec:Intro}

Dark matter haloes grow through repeated, hierarchical mergers, and simulations of this process predict that the central regions of merging haloes can survive as self-bound substructures within the final system. These subhaloes should host most of the visible galaxies in the low-redshift Universe, and may be one of the best environments in which to explore and constrain the particle nature and non-gravitational properties of dark matter \citep[see e.g.][for a review]{bullock2017}. 
 
In general, observational tests of this predicted dark matter distribution include galaxy dynamics, gravitational lensing, and `indirect detection' (i.e., searches for radiation or products from dark matter annihilation). All of these tests are most sensitive where dark matter densities are highest, at small radii within haloes and subhaloes. The annihilation signal in particular will depend sensitively on subhalo density profiles, concentration, mass loss, and disruption \cite[as summarized in, e.g.][]{okoli2018,ando2019}. Since this signal is proportional to the local density squared, one must understand the details of the density distribution within haloes and subhaloes, down to the smallest scales on which CDM can cluster, to place reliable constraints on dark matter particle properties.

The visible structure of galaxies can be used to trace the central density distribution in haloes \citep[e.g.][and references therein]{taylor2019}, but only out to a few per cent of the virial radius \citep{kravtsov2013, somerville2018}. Typically, measurements of the inner core or cusp of dwarf-galaxy haloes require tens of thousands of stars and are sensitive to dynamical assumptions \citep{chang2021}. 
Gravitational lensing studies can probe the total mass distributions around galaxies, groups, and clusters more directly. However, while gravitational lensing has the advantage of being insensitive to the dynamical state of the system, the effect is normally weak enough that good models of the lens potential are required. As first shown by \cite{natarajan2002} and since measured in several studies \cite[e.g.][]{li2016, niemiec2017, sifon2018, dvornik2020, kumar2022, wang2024}, the tidal truncation of galaxy haloes within clusters relative to those of field galaxies is detectable in lensing and needs to be included in models of the total mass distribution to obtain accurate halo mass estimates \citep{baltz2009}.

As subhaloes orbit within a larger system post-merger, they will lose mass through tidal stripping. This process generally works from the outside in \citep[e.g.][]{hayashi2003,diemand2007}. However, the exact effect of tidal stripping in the innermost part of the subhalo has remained unclear from previous work, given the resolution limits of simulations. In principle, repeated mass loss can also disrupt substructure completely, as is often seen in cosmological simulations of halo formation. Recent work suggests that much of this disruption is artificial, and due to insufficient resolution \citep{vandenbosch2017, vandenbosch2018,vandenbosch2018b,benson2022}, and that the central density of cuspy subhaloes should be preserved to arbitrarily small masses \citep{kazantzidis2006, errani2020, amorisco2021, drakos2022}. 

Fundamentally, subhalo structure depends on the initial properties of the halo at infall and the subsequent effect of tidal evolution.  Studies of the earliest forming haloes---which are expected to exist as subhaloes at low 
redshifts---suggest that these objects are cuspier than systems that form later, and have an inner density profile of the form $ r^{-1.5}$ \citep[e.g.][]{ishiyama2014,angulo2017,ogiya2018,delos2019a,delos2023}\footnote{This is steeper than the commonly-used NFW profile, which has an inner density profile of $ r^{-1}$, as discussed in Section~\ref{sec:halo_models}.} Given the strong dependence of the annihilation signal on the dark matter density on the smallest scales and in the densest systems, this has enormous implications for current annihilation constraints \citep{delos2023b}. For instance, it has been suggested that there should be $\sim 10^{16}$ ``prompt cusps" in the Milky Way \citep{delos2022}, which would contribute 20-80 per cent of a putative annihilation $\gamma$-ray background \citep{stucker2023}.

The current study investigates how different modelling assumptions for subhalo density profiles impact dark matter annihilation and lensing signals in individual subhaloes. In particular, we focus on the consequences of assuming that the central density of cuspy subhaloes is preserved during tidal evolution. The structure of this paper is as follows: first, in Section~\ref{sec:model}, we summarize the subhalo models considered in this work.  In Section~\ref{sec:concentration}, we show how the concentration evolves in these different models. In Sections~\ref{sec:anni} and \ref{sec:lens},  we explore the implications of a conserved central density to the dark matter annihilation rate and lensing signal, respectively. Finally, we discuss our results in Section~\ref{sec:discuss} and conclude in Section~\ref{sec:conc}.
	
\section{Halo Models} \label{sec:model}

\subsection{Isolated dark matter haloes} \label{sec:halo_models}

Most of our understanding of halo structure comes from cosmological simulations. One of the most important results from these studies is that isolated haloes have a universal density profile (UDP) when averaged spherically. The UDP was originally approximated in \cite{navarro1996,navarro1997}---hereafter NFW---as:
\begin{equation}
\rho_{NFW}(r) = \dfrac{\rho_0 r_{\rm s}^3}{r(r+r_{\rm s})^2} \,\,\, ,
\end{equation}
where $\rho_0$ is a characteristic density and $r_{\rm s}$ is the scale radius, describing the point where the logarithmic slope is ${\rm d} \log \rho/ {\rm d} \log r = -2$. 

Though NFW profiles are  still commonly assumed in cosmological calculations, dark matter haloes in cosmological simulations are better described by an Einasto profile \citep{einasto1965}
\begin{equation}
\rho_{Ein}(r) = \rho_{-2} \exp \left( -\dfrac{2}{\alpha} \left[ \left( \dfrac{r}{r_{-2}}\right)^\alpha -1\right]\right)  \,\,\, ,
\end{equation}
where $\rho_{-2}$ and $r_{-2}$ correspond to  where the logarithmic slope is ${\rm d} \log \rho/ {\rm d} \log r = -2$. Though this fit provides a better description of the UDP down to the resolution limit of simulations \citep{navarro2004,gao2008,klypin2016}, the detailed form of the profile at very small radii remains uncertain. 
 
 The ``shape'' parameter, $\alpha$, controls the inner slope of the density profile, with small $\alpha$ values corresponding to ``cuspier" centres. The mean value of $\alpha$ increases with peak height within the cosmological density field \citep{gao2008,klypin2016};
for low peaks (i.e. low-mass and/or low redshift haloes) with values of $\alpha \sim$  0.2--0.25, the Einasto profile is very similar to the NFW profile, while for rare, recently-formed and/or high-mass objects, it is smoother with a shallower inner slope. Aside from their density profiles, haloes are often described by their mass and concentration. We discuss the definition of the evolution of the concentration parameter in Section~\ref{sec:concentration}.

\subsection{Tidally-stripped dark matter subhaloes}

As haloes merge hierarchically, tidal stripping primarily removes material from the outer radii, causing characteristic changes to the density profile. These tidally stripped systems are often described by empirical models calibrated to simulations. The first of these models was developed in \cite{hayashi2003}---hereafter H03---which posits that at large radii, the slope of tidally stripped systems is ${\rm d} \log \rho/ {\rm d} \log r = -4$, and the central density is decreased, according to the parameterized equation:
\begin{equation}  \label{eq:H03}
\rho(r) = \dfrac{f_t}{1 + (r/r_{te})^3} \rho_{NFW} (r) \,\,\,.
\end{equation}
 where $r_{te}$ is an ``effective" tidal radius, and $f_t$ describes the reduction in central density. Both of these parameters can be estimated using a single parameter---the bound mass fraction, $f_{\rm b}$, of the satellite:
	\begin{equation}
	\begin{aligned}
	\log(r_{te}/r_{\rm s}) &= 1.02 + 1.38 \log f_{\rm b} + 0.37 (\log f_{\rm b})^2\\
	\log f_t &= -0.007 + 0.35 \log f_{\rm b} \\
	& \hspace{1.5cm} +0.39 (\log f_{\rm b})^2 + 0.23 (\log f_{\rm b})^3 \,\,\,.
	\end{aligned}
	\end{equation}
In H03, the bound mass fraction $f_{\rm b}$ is defined as the mass of the bound satellite compared to the mass of an untruncated NFW profile within radius $r_{\rm vir}=10\,r_s$.

The disadvantage of empirical methods like H03 is that these models are only valid across the range of cases considered in the simulations and are, therefore, generally limited to specific density profiles and orbital parameters. Additionally, these models may include numerical artefacts. For example, H03 predicts a reduction in the central density, which is a consequence of an artefact of the approximation they used to set up the initial conditions in their simulations \citep{kazantzidis2004}. More recent empirical models \cite[e.g.][]{penarrubia2010, green2019} predict similar trends but with less reduction in the central density. In general, empirical models accurately reproduce a suite of isolated simulation results by construction, but cannot be reliably extrapolated beyond the suite of simulations. Since the very centre of haloes will never be satisfactorily resolved in $N$-body simulations due to two-body interactions, these parameterizations are not able to predict the evolution at small radii/small mass fractions \citep{drakos2022}.
	
An alternative approach is to model the evolution of subhaloes using physical principles. This approach is generally less accurate than calibrated empirical models, with the possible exception being the Energy Truncation model developed in \cite{drakos2017,drakos2020}; see also \citep[][]{widrow2005}. The Energy Truncation model is based on the finding that particles are primarily stripped as a function of their energy \citep[e.g.][]{choi2009, drakos2017,stucker2021}; particles that are less gravitationally bound get stripped first, regardless of their radius.

In practice, this energy truncation is performed by lowering and shifting the distribution function, $f_0$, of the initial profile, according to 
\begin{equation}
f(\mathcal{E}) = f_0(\mathcal{E}+\mathcal{E}_T) -  f_0(\mathcal{E}_T )\,\,\, ,
\end{equation}
where $\mathcal{E} =\Psi(r) - v^2/2$ is the relative ``binding" energy and $\Psi(r)=-\phi(r)$ is the relative potential energy. The parameter $\mathcal{E}_T$ is termed the truncation energy and sets the mass and tidal radius of the truncated system. Then, the potential of this system can be found by solving Poisson's equation and  Eddington's inversion of the density profile \citep{eddington1916}. This procedure is analogous to the derivation of the King model \citep{king1966} and has been described in detail in \cite[][]{drakos2017,drakos2020, drakos2022}. The free parameter $\mathcal{E}_T$ can equivalently be expressed as a bound mass or tidal radius. In this work we primarily express subhaloes by their bound mass, but, in principle,  $\mathcal{E}_T$ can be determined from the orbital parameters of the merger \citep{drakos2020}, and thus, the Energy Truncation model has no free parameters (i.e., it does not need to be tuned to simulations). 

\begin{figure}
	\includegraphics[width=1\textwidth]{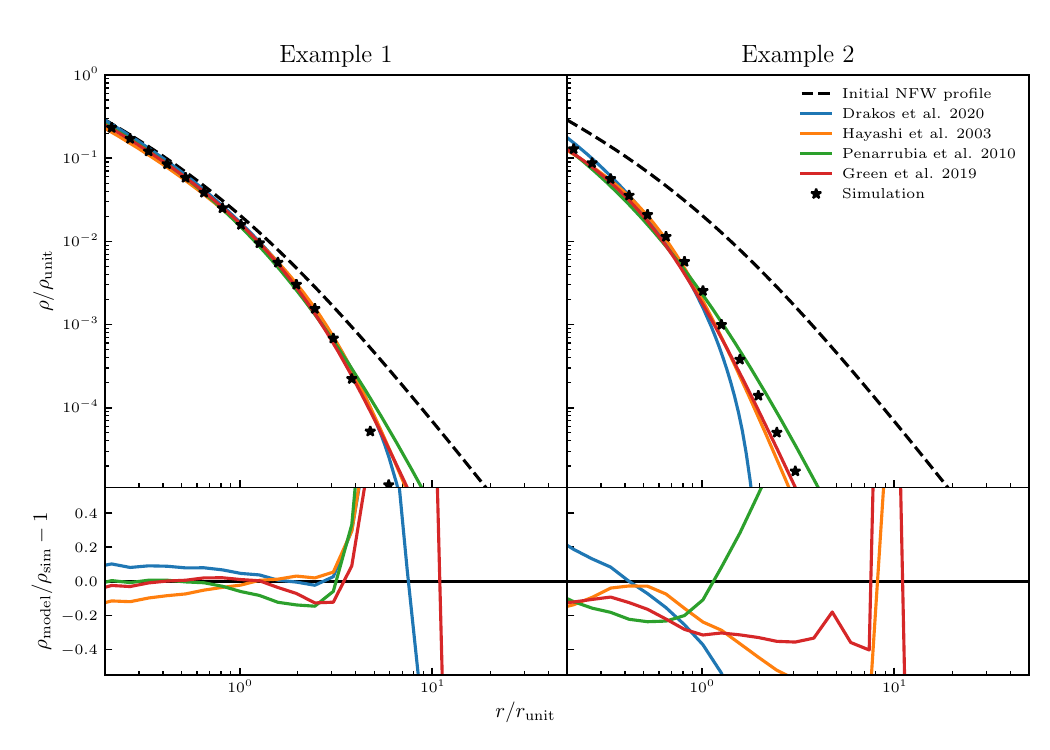}
	\caption{Comparison of models for tidally stripped NFW haloes. The simulation results are from an idealized simulation of an NFW subhalo inside a fixed background potential after 5 orbits. We have used two simulations; Example 1 is demonstrative of a case in which the Energy Truncation Model works quite well, and Example 2 is a case where the Energy Truncation Model performs less accurately; these correspond to the Slow and Fast Mass loss simulations in \citep{drakos2022}. The Energy Truncation Model \citep{drakos2020} has no free parameters, while the three parametric models \citep{hayashi2003, penarrubia2010, green2019} are calculated from the bound mass of the simulated subhalo. Overall, all of the models do a comparable job of predicting the subhalo density profile. }
	\label{fig:Compare_sims}
\end{figure}

Figure~\ref{fig:Compare_sims} demonstrates how well the Energy Truncation model predicts the subhalo density profile evolution compared to empirical models calibrated to match simulations  \citep{hayashi2003,penarrubia2010,green2019}. While empirical models are calibrated to match simulation results,  the impact of numerical artefacts on the central density profile remains uncertain. The Energy Truncation model predicts slightly higher central densities than the empirical models but agrees well with isolated simulations and provides comparable accuracy to parameterized models within the radii that are well resolved in simulations \citep{drakos2020}. Additionally, this model has been shown to reproduce not only the density profile but also the phase-space structure of tidally stripped systems undergoing gradual mass loss, particularly when evaluated near apocentre \citep{drakos2017}. However, we do not expect the Energy Truncation Model to accurately describe all scenarios---for instance, it may not perform well in systems experiencing rapid or impulsive stripping (e.g., disk shocking), where energy redistribution can be more complex.

Since the Energy Truncation model predicts that the central density is preserved---unlike most empirical parameterizations---we adopt this model as a physically motivated upper bound on central densities. Additionally, because the Energy Truncation model is applicable to any tidally stripped collisionless system \citep{drakos2022}, it offers a flexible approach for studying different profile models without requiring calibration to simulations.

\subsection{This work}\label{sec:thiswork}

Overall there is some fundamental uncertainty in the original density profile of isolated haloes (e.g. NFW  and the range of Einasto profiles) and the way they are modified by tidal stripping.  It is impossible to completely resolve these uncertainties by direct numerical simulation: given the limited dynamic range of simulations, we will never be able to resolve the very centre of low-redshift haloes in any near future. As a result, the central density of subhaloes may be higher than previously expected. In the remaining sections of the paper, we examine how conserved central density and varying initial profiles influence the dark matter annihilation and galaxy lensing signals.  

\begin{figure} 
    \centering
	\includegraphics[]{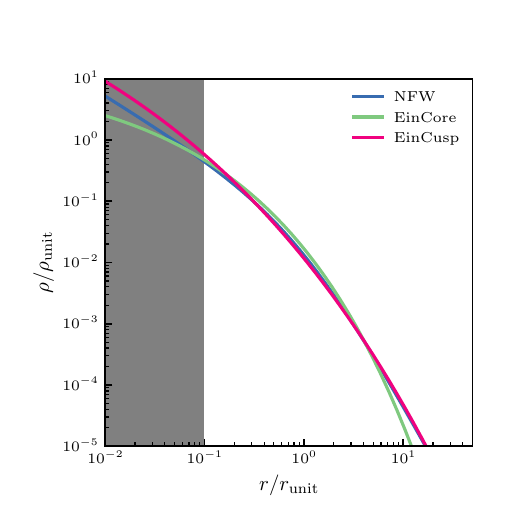}
	\caption{Comparison of initial, unstripped profiles used in this work. All of these are normalized to have the same virial mass and a concentration of $c=10$. At radii smaller than approximately 10 per cent of the scale radius (grey box)---which are beyond what is typically resolved in isolated simulations---there is a significant difference in the central density of the different models. }
	\label{fig:Profs_Init}
\end{figure}

We will consider three initial profile models
\begin{enumerate}
\item an NFW profile, for comparison with previous work
\item a more ``cuspy" Einasto profile (EinCusp\footnote{In \citep{drakos2022}, EinCusp and EinCore were called EinLow and EinHigh, respectively.}; $\alpha=0.15$), and 
\item a more ``cored" Einasto profile (EinCore; $\alpha=0.3$).
\end{enumerate}
These Einasto parameters were chosen since they span the range of parameters found in simulations \citep[e.g.][]{gao2008,klypin2016}.  The lower value is typical of low peaks, i.e. late-forming or low-mass objects, while the high value is typical of early-forming or high-mass objects. For comparison, a commonly adopted fiducial Einasto parameter is 0.17 \citep[e.g.][]{lazar2023}. The NFW profile should be intermediate between the two, and most similar to an Einasto profile with $\alpha \sim 0.2$--0.25. We will use units $G=1$, $r_{\rm unit} = r_{-2}$, and $M_{\rm unit} = M_{\rm NFW}(r<10\,r_{-2})$; that is, the profiles will be normalized to have the same virial mass and a concentration of $c=10$. The density unit is then $\rho_{\rm unit} = M_{\rm unit} r_{\rm unit}^{-3}$. As shown in Figure~\ref{fig:Profs_Init}, although these three profiles have the same virial mass and concentration, at radii less than 10 per cent of scale radius, the central densities begin to differ considerably. 

We will primarily use the Energy Truncation model to describe our tidally stripped systems and compare these results to two commonly used approximations in the literature: (1) the H03 model and (2) a sharp truncation (ST) of an NFW model. The H03 model is known to underestimate the central density, while the ST model preserves the NFW profile precisely within the tidal radius; therefore, the Energy Truncation model can be used as an intermediate, physically-grounded alternative to both the H03 (which suffers from numerical limitations) and the sharp truncation model (which does not account for any internal structural change). The parameterizations of \cite{penarrubia2010} and \cite{green2019}  are expected to fall somewhere between the Energy Truncation model and the H03 model. Given this, we adopt the Energy Truncation model as a conservative upper bound on subhalo density for cuspy systems, representing a scenario in which the central density is fully preserved.

\begin{figure}
	\includegraphics[width=1\textwidth]{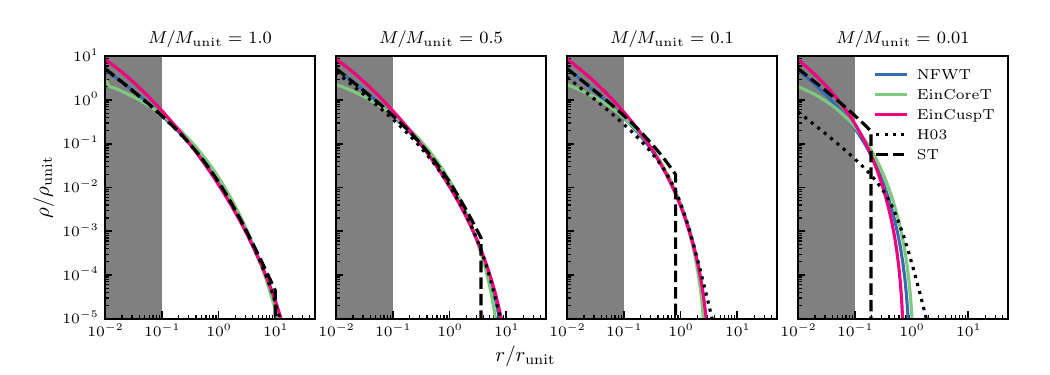}
	\caption{Tidally stripped subhalo density profiles. The models are summarized in Table~\ref{tab:ICs}. Each panel is stripped to a different bound mass. As in Figure~\ref{fig:Profs_Init}, the grey box indicates radii that are beyond what is typically resolved in isolated simulations. The Energy Truncation model predictions (NFWT, EinCoreT, and EinCuspT) are similar, except at very small radii. The two other stripping models differ significantly from the Energy Truncation model. The H03 model is known to underestimate the density of the subhaloes at small mass fractions, while the ST model only agrees with NFWT at small radii.}
	\label{fig:ProfileCompare}
\end{figure}

A summary of the models for describing the tidally stripped systems is given in Table~\ref{tab:ICs}, and Figure~\ref{fig:ProfileCompare}. These models will allow us to compare 
\begin{enumerate}
\item Differences in predictions caused by assumptions about the initial profile model under the hypothesis of a conserved central density; i.e., between NFWT, EinCoreT, and EinCuspT
\item Differences in predictions caused by assumptions of how a tidally stripped profile evolves, i.e., between NFWT, H03, and ST.
\end{enumerate}

\begin{table*} 
\centering
	\caption{ \label{tab:ICs}Summary of the subhalo models used in this work. For all models, subhaloes are assumed to have an initial infall mass of  $M_{\rm unit} = M_{\rm NFW}(r<10\,r_{-2})$ and a concentration of $c=10$.  This leaves one free parameter; the bound mass, $M$, of the satellite. }
	\begin{tabular}{|c |c| c|}
		\hline
		Profile Name & Initial Profile &  Stripping Model \\ 
		\hline
		NFWT & NFW &  Energy Truncation\\
		EinCoreT   & EinCore &  Energy Truncation \\
		EinCuspT & EinCusp &  Energy Truncation  \\
		ST & NFW & Sharp truncation \\
		H03 & NFW &  Parametric model from \citep{hayashi2003}\\
		\hline
	\end{tabular}	
\end{table*}

\section{Evolution of the concentration parameter} \label{sec:concentration}

Density profiles can be characterized by a concentration parameter, which reflects the distribution of mass within the system. 
For isolated haloes, the concentration parameter $c$ is traditionally defined in terms of the radius at which the logarithmic density profile has a slope of $-2$, i.e., $c= r_{\rm vir}/r_{-2}$ \citep{navarro1997}, where $r_{\rm vir}$ is the virial radius. 
As discussed in \cite{klypin2016,drakos2019b}, this definition does not capture deviations from an NFW profile. However, concentration is broadly correlated with formation epoch, early-forming haloes being more concentrated, though the details of the relationship are complicated \citep[e.g.][]{wechsler2002,zhao2003, zhao2009,wong2012,ludlow2014,correa2015,wang2020}. 

In cosmological simulations, the virial radius is normally defined in terms of a (redshift-dependent) mean enclosed density. In idealized merger studies, the definition of the virial radius is somewhat arbitrary, but as in Section~\ref{sec:thiswork}, we assume the satellite haloes begin merging with a concentration of $c=10$. This corresponds to an enclosed density of $\bar{\rho}_{\rm vir}=0.2387 \, \rho_{\rm unit}$.
As the satellite is stripped, we solve for $r_{\rm vir}$ as the radius where $\bar{\rho} (r_{\rm vir})=\bar{\rho}_{\rm vir}$. The scale radius, $r_{-2}$, is calculated directly by numerically differentiating and then solving  ${\rm d}\log_{10} \rho / {\rm d} \log_{10} r =-2$. For the ST profile, once the tidal radius (defined below) is within the original scale radius, we use $r_{-2} = r_t$.

As we will show below, this first definition of concentration generally {\it increases} as mass is lost since the scale radius shrinks faster than $r_{\rm vir}$.  
While this definition is useful for tracking how the bound remnant of a subhalo increases in mean density as it loses its lower-density outer material, $r_{\rm vir}$ has no clear physical significance once a halo has become a subhalo since its density no longer tracks the original cosmological mean value. Thus, in addition to the classical concentration, $c=r_{\rm vir}/r_{-2}$, for stripped systems we will consider an ``effective" concentration parameter, $c_t = r_t/r_{-2}$. As argued in \cite{bartels2015,okoli2018}, this definition may be more natural as it better reflects the higher mean density of subhaloes and the lower concentration the subhalo had when it was originally accreted. In general, we define the tidal radius $r_t$ to be the radius at which the density profile is zero.  With this definition, the H03 profile does not have a tidal radius, so for this profile, we will instead use the effective tidal radius,  $r_{\rm te}$, as defined in Equation~\eqref{eq:H03}.

\begin{figure}
    \centering
	\includegraphics[]{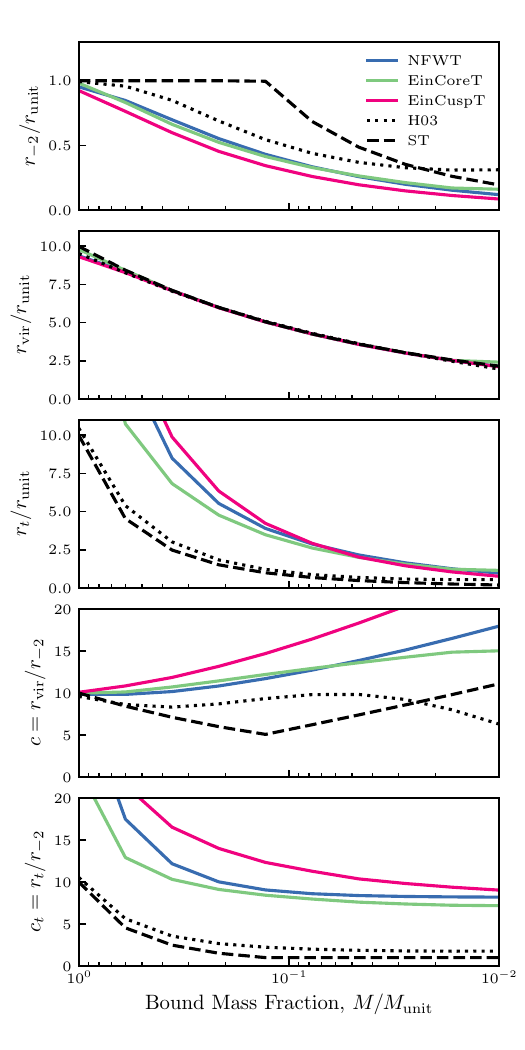}
	\caption{Evolution of the parameters used to characterize subhalo concentration as a function of bound mass fraction, for various profiles: (1) the scale radius $r_{-2}$, (2) the virial radius $r_{\rm vir}$; the radius enclosing an overdensity of $\bar{\rho}_{\rm vir}=0.2387 \, \rho_{\rm unit}$, (3)  the tidal radius,  (4) the concentration parameter $c=r_{\rm vir}/r_{-2}$, and  (5) the effective concentration $c_t = r_t/r_{-2}$. All three Energy Truncation models evolve similarly to each other, with the EinCuspT concentration being about 20 per cent higher than the NFWT, and EinCoreT models. The concentration parameters calculated from the other two models, H03 and ST, evolve quite differently. }
	\label{fig:Conc}
\end{figure}

With these definitions, Figure~\ref{fig:Conc} shows the evolution of the scale radius, the virial radius, the tidal radius, and the two concentration parameters as  the bound mass fraction decreases from left to right on the $x$ axis. For all three Energy Truncation models (NFWT, EinCoreT and EinCuspT) the scale radius decreases monotonically as the halo is tidally stripped, as expected \citep{drakos2022}. The relative change in $r_{-2}$ is always slower than the relative mass loss rate, however, such that when the system has lost 90\%\  of its mass, $r_{-2}$ has only decreased to 20--50 per cent of its initial value. The behaviour for the evolution of $r_{-2}$ in the H03 approximation is similar, although this model generally predicts a scale radius that is 10--50 per cent higher than the Energy Truncation model. The ST approximation predicts a constant $r_{-2}$ until the tidal radius is less than the scale radius (at approximately $0.5\,M_{\rm unit}$), at which point it rapidly decreases.

The virial radius shows a similar monotonic decrease that is approximately the same for all three Energy Truncation models; it varies only because of the slightly different relationship between mean density versus radius for each profile. Once again, $r_{\rm vir}$ changes more slowly relative to its initial value than the bound mass. 

The tidal radius, $r_t$, also decreases with mass, however---unlike $r_{\rm vir}$---the value of $r_t$ varies widely between the models. For the three Energy Truncation models, $r_t$ is initially very large and decreases rapidly until the mass is approximately 10 per cent of $M_{\rm unit}$. At smaller masses, the three Energy Truncation models agree quite well. The H03 and ST approximations do not agree with Energy Truncation models and, instead, look similar to each other. Both of these approximations start with $r_t=r_{\rm vir}$ and then rapidly decrease. We emphasize that the tidal radius is defined differently for the H03 model, using the parameter $r_{te}$. This effective tidal radius is used in the H03 parameterization but does not have a clear physical meaning. We also note that the ``virial'' radius is typically larger than the tidal radius, except for the Energy Truncation models at high mass fractions; this means the virial radius is simply given by $r_{\rm vir} = (4  \pi \bar{\rho}_{\rm vir } / 3 M)^{1/3}$.

Taking the ratio of the two outer radii to the scale radii, we calculate the corresponding concentration parameters. For the Energy Truncation models, $c=r_{ r_{\rm vir}}/r_{-2}$, increases with decreased mass. We emphasize, however, that this apparent increase in $c$ comes from the definition of the virial radius, which tracks the original density of the system and will extend beyond the tidal radius. Physically, the material still bound to the subhalo within its tidal radius is less concentrated---this is better captured by the second concentration parameter, $c_t$. We show how the second effective concentration parameter varies with mass loss in the bottom panel of  Figure~\ref{fig:Conc}. Once the subhaloes have been stripped to about 20\% of their original mass, the effective concentration is approximately constant at $c_t\approx10$. Given the faster decline in its scale radius, the EinCuspT profile has higher concentrations than the NFWT and EinCuspT profiles. The ST and H03 models predict a lower $c_t$ by an order of magnitude or more compared to the Energy Truncation model; though in the case of the H03, this is largely because the tidal radius definition is different.

Overall, all profiles and mass loss models show certain broad trends. If we take the mass of field haloes to be their virial mass within $r_{\rm vir}$, and the mass of stripped subhaloes to be the bound mass within the tidal radius $r_t$, then stripped subhaloes have higher mean densities than field haloes of the same mass, but their bound material is more uniformly distributed than that of field haloes. Thus, concentrations relative to the original virial boundary generally increase, while concentrations relative to the tidal radius decrease. Nonetheless, significant differences in concentration evolution are predicted simply based on differences in (1) the initial profile model and (2) the tidal stripping model. 
Finally, we note that the Energy Truncation model predicts that at low bound mass fractions, the scale radius and tidal radius will evolve at similar rates, resulting in a constant $c_t$ value.

\section{Dark matter annihilation} \label{sec:anni}

One potential technique for determining the identity of the dark matter particle is through dark matter annihilation. To place constraints on particle masses and interaction cross-sections, it is important to have accurate predictions of the distribution of dark matter within subhaloes since the dark matter annihilation signal depends sensitively on substructure. A common approach to modelling the contribution of substructure to annihilation signals is to draw subhalo properties such as infall redshift, mass, concentration, and/or orbital properties from random distributions or known scaling relations, and then to integrate the contribution from each subhalo to the total signal \citep[e.g.][]{bartels2015, han2016, stref2017,okoli2018,hiroshima2018,hutten2019,ibarra2019, delos2019d, stref2019,facchinetti2022}.

Assuming the dark matter particles are Majorana WIMPs (weakly interacting massive particles) that can annihilate with one another, the rate at which dark matter annihilates is given by:
\begin{equation}
R(V)=\dfrac{\langle\sigma v \rangle}{2 m^2} \int_V \rho^2 dV' \,\,\, ,
\end{equation}
where $\langle\sigma v \rangle$ is the velocity-averaged annihilation cross section, $m$ is the mass of the dark matter particle, $\rho$ is the density and $V$ is a volume \citep[e.g.][]{taylor2003}. Thus, the annihilation rate is proportional to the quantity
\begin{equation}\label{eq:J}
J(V) \equiv \int_V \rho^2 {\rm d}V' \,\,\, ,
\end{equation}
 which is sometimes called the $J$-factor\footnote{ Though the $J$-factor is traditionally expressed as the flux along the line-of-sight \citep{bergstrom1998}, here we follow the definition of \cite{delos2019b}, where the $J$-factor reflects the absolute annihilation luminosity. }, and has dimensions mass$\times$density. This quantity is sensitive to the high-density inner region of the halo and does not depend strongly on the choice of integration volume, as long as  that volume is large enough to enclose the region of high density.

Since the $J$-factor depends on the overall size and mass of the system, it is also convenient to define a dimensionless quantity,
\begin{equation}\label{eq:B}
f_M(V) = \dfrac{1}{\bar{\rho}^2 V} \int_V \rho^2 dV' \,\,\, ,
\end{equation}
where $\bar{\rho}$ is the mean density within the volume $V$. We will refer to this as the ``flux multiplier'', as in \cite{taylor2003}.\footnote{This quantity has also been called the halo boost factor \citep{okoli2018}}. Unlike the annihilation rate, this quantity is sensitive to the integration volume. In the case of CDM haloes, it is common to use the spherical volume within the virial radius. 

In terms of the flux multiplier, the annihilation rate for a single halo can be expressed as:
\begin{equation}
R(V)=\dfrac{\langle\sigma v \rangle}{2 m^2} \bar{\rho}^2 V f_M(V) \,\,\, ,
\end{equation}
where the first term, $\langle\sigma v \rangle / 2 m^2$, depends on the particle physics of the dark matter candidate, the second term $\bar{\rho}^2 V$  depends on the mean density and volume  (or equivalently the size and total mass) of the system, and $f_M(V)$ characterizes the mass profile of the halo within the volume $V$.

\begin{figure}
 	\includegraphics[width=1\textwidth]{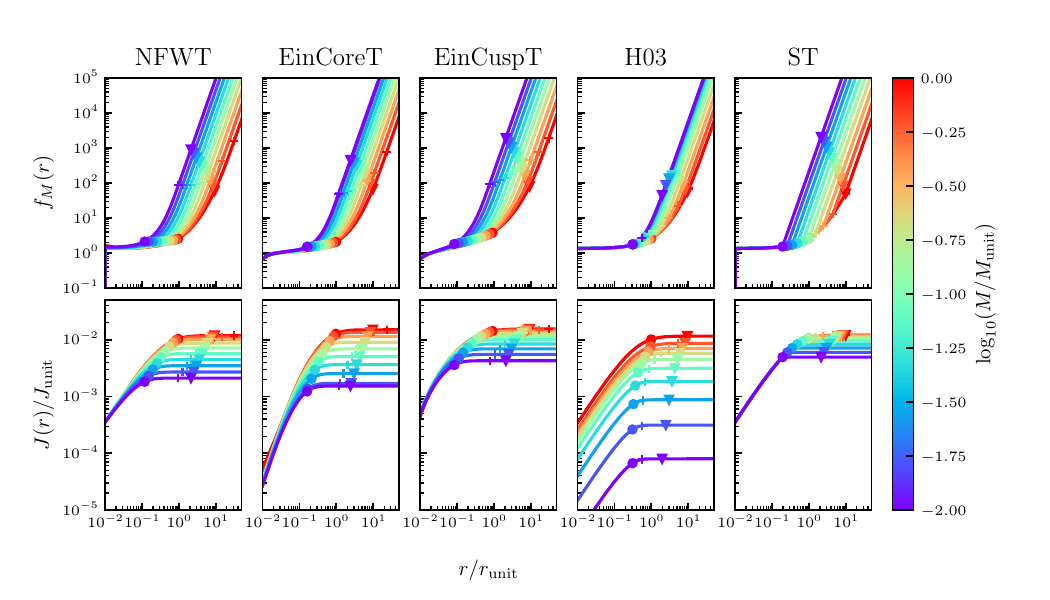}
 	\caption{The flux multiplier, $f_M(r)$ (top) and $J$-factor, $J(r)$ (bottom) for tidally stripped subhalo models, as a function of the radius of the spherical volume within which the integral is calculated. The colour scale indicates the degree of stripping. The characteristic radii $r_{-2}$, $r_{\rm vir}$ and $r_t$ are labelled as circles, triangles, and crosses, respectively. The flux multiplier is approximately constant within the scale radius and then increases as $r^3$. Conversely, the $J$-factors increase rapidly within the scale radius and then are constant at large radii. This plot demonstrates how the flux multiplier and $J$-factor measurements depend on the choice of integration volume.}
 	\label{fig:Annihil}
 \end{figure}

The flux multiplier ($f_M$; top) and $J$-factor (J; bottom)
 calculated within a spherical volume of radius $r$ are shown in Fig~\ref{fig:Annihil} for each profile model and various degrees of mass loss. For the flux multiplier (which is a measurement of how inhomogeneous the mass distribution is), the parameter is roughly constant within the scale radius, $r_{-2}$, and then increases rapidly with radius (with $f_M \propto V$ outside the tidal radius). Inside $r_{-2}$, there is little evolution in the flux multiplier as mass is stripped, while outside this radius, the flux multiplier is higher for more stripped systems. Conversely, the $J$-factor increases with a radius for $r<r_{-2}$ and then is approximately constant at larger radii, but its value decreases as the system is stripped, especially in model H03.

\subsection{Flux multiplier evolution}
 
\begin{figure}
	\includegraphics[width=1\textwidth]{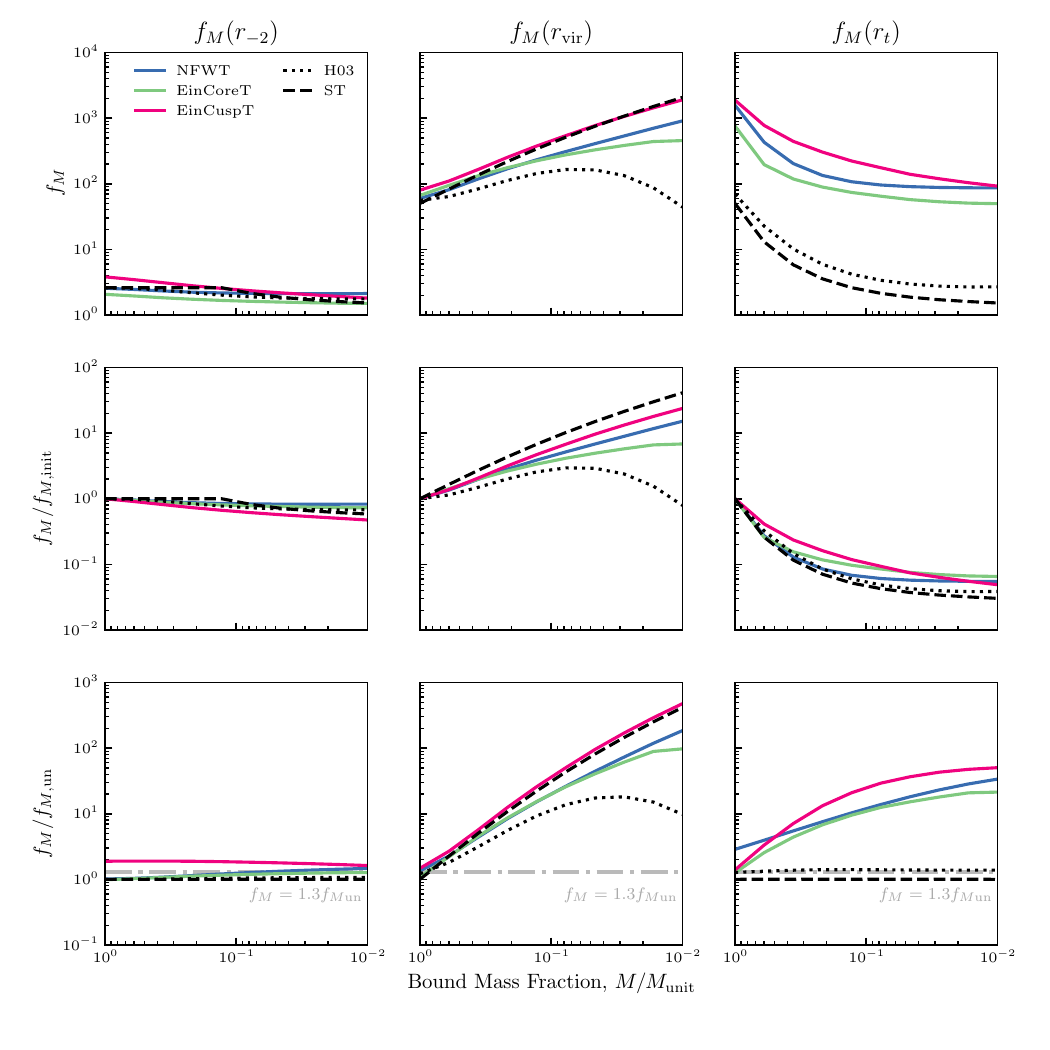}
	\caption{Evolution of the flux multiplier  calculated within different volumes. The top row shows the flux multiplier , the second panel shows the evolution from the initial value, and the bottom shows the ratio between the flux multiplier of the tidally-stripped profiles and the initial, untruncated profiles. }
	\label{fig:Boost}
\end{figure}

In the top row of Figure~\ref{fig:Boost}, we show how the flux multiplier evolves as a function of bound mass using different volumes. As expected from Figure~\ref{fig:Annihil}, when calculated within the scale radius, the flux multiplier stays roughly constant. Within the virial radius---except the H03 approximation---the boost factor increases as mass is stripped. Finally, within the tidal radius, the flux multiplier \emph{decreases} as the profile is stripped, with both the H03 and ST approximation having a lower flux multiplier than the Energy Truncation Model by two orders of magnitude. This difference is largely because of the different volumes in which the flux multiplier is calculated. If we focus on the evolution of the flux multiplier relative to its initial value (second row), this difference disappears.

It is also worth comparing predictions of flux multipliers to the values obtained using the untruncated profile.  The bottom row of Figure~\ref{fig:Boost} shows the ratio $f_M/f_{M, {\rm un}}$, where $f_{M, {\rm un}}$ is the initial un-truncated profile. We compare our findings to the constant value of 1.3, since \cite{taylor2003} showed that the H03 model follows the relation $f_M \simeq1.3\, f_{M, {\rm un}}$ when measured within $r_{te}$. 

When measured inside the scale radius, the flux multiplier  is constant for all models and consistent with $f_M=1.3 f_{M, {\rm un}}$. When measured within the virial radius, we find much higher values of $f_M/f_{M, {\rm un}}$. This result is because the virial radius is typically larger than the tidal radius in these systems. Since mass loss reduces the density in the outer parts of the stripped halo,  this makes the density distribution of stripped systems more centrally concentrated and less homogenous, and thus it increases $f_M$. 

When measured within the tidal radius, we find the ST and H03 approximations agree with the $f_M\simeq1.3\,f_{M, {\rm un}}$ prediction, as expected. The more accurate Energy Truncation stripping models predict a flux multiplier  that is more than 10 times higher by the time the bound mass fraction is less than 10 per cent. We conclude that the flux multiplier  is extremely sensitive to the details of mass loss within the tidal radius, and may have been underestimated in previous work. 

\subsection{$J$-factor evolution}

\begin{figure}
    \centering
 	\includegraphics[]{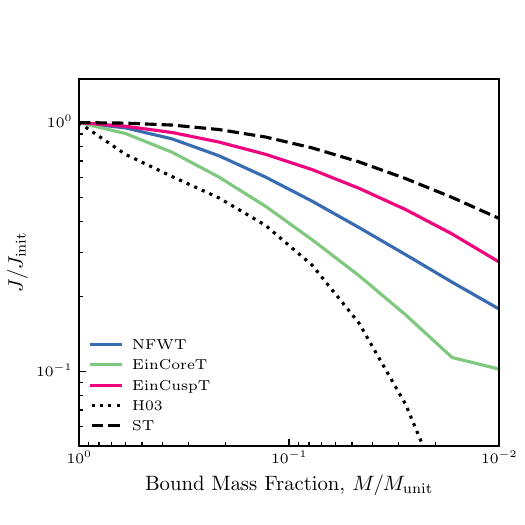}
 	\caption{Change in the $J$-factor versus bound mass fraction predicted from our models. There are large differences between all of the models, suggesting that $J$-factor predictions are very sensitive to assumptions about subhalo profiles. Further, the $J$-factor is not constant as is often assumed but can decrease significantly by the time the subhalo is stripped to half of its mass.}
 	\label{fig:Jfactor}
 \end{figure}

Ultimately we are interested in predicting the total annihilation rate in dark matter subhaloes, which is proportional to the $J$-factor defined in Equation~\ref{eq:J}.  In Figure~\ref{fig:Jfactor}, we show predictions for how the $J$-factor evolves with mass loss for the different profile models. We calculate the $J$-factor within the tidal radius but note that the $J$-factor is insensitive to this choice. As with the flux multiplier , we find large differences in the $J$-factor between the different models. By the time the profile is stripped to 5 per cent of the initial mass, there is a factor of 5 difference in the annihilation signal between the ST approximation and the H03 approximation, while the Energy Truncation predictions lie between the two. It is often assumed that since the high-density inner regions of subhaloes are resilient against tidal forces, the annihilation rate will remain fairly constant during tidal stripping \citep{bartels2015}. Figure~\ref{fig:Jfactor} shows that this assumption is incorrect, once mass loss reaches a significant level.

 Comparing the three Energy Truncation models, the $J$-factor in the EinCuspT case (whose density profile has a steeper central cusp; Figure~\ref{fig:Profs_Init})  decreases the least, while the value from the flatter EinCore profiles decreases the most.  The ST model (dashed line) is expected to always overestimate the annihilation signal. Although the H03 model (dotted line) gives a reasonable match to the stripped density profile except at very low masses/radii (Figure~\ref{fig:ProfileCompare}), this model predicts much lower annihilation rates overall. 

\begin{figure}
    \centering
  	\includegraphics[]{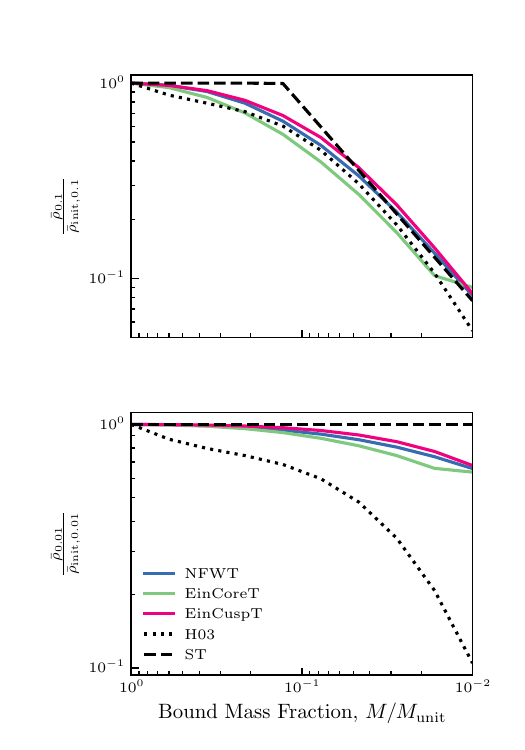}
  	\caption{Change in the central density versus bound mass fraction for various profiles. The enclosed density is calculated in either 10 per cent (top) or 1 per cent (bottom) of the initial virial radius. The enclosed density decreases over time, with slight differences between the three Energy Truncation profiles. For the ST approximation, the enclosed density is constant over time within the tidal radius and thus over-predicts the more accurate NFWT model. The H03 approximation---which is known to underestimate the central density---predicts a significantly larger decrease in the central density compared to the NFW model.}  
  	\label{fig:CentralDens}
  \end{figure}

 The variation in the predicted $J$-factor between models reflects the sensitivity of the annihilation signal to the central density. In Figure~\ref{fig:CentralDens} we show how the enclosed central density evolves with bound mass. We consider the mean density within 10 per cent ($\bar{\rho}_{0.1}$; top panel) and 1 per cent ($\bar{\rho}_{0.01}$; bottom panel) of the initial virial radius (as described in Section~\ref{sec:concentration}). The cuspy EinCusp profile has the smallest change in central density, which is why the $J$-factor for this model changes slowly. When comparing approximations for tidally stripped NFW profiles, the central density of the ST model is constant as expected, as long as the tidal radius exceeds the radius used to calculate the enclosed density. Relative to the NFW results, both the ST and the H03 model predictions agree closely with the NFWT model for the enclosed density within 10 per cent of the virial radius. However, the ST approximation over-predicts the central density within 1 per cent of the virial radius, while the H03 predicts a signficantly lower enclosed density, as discussed in Section~\ref{sec:thiswork}.

\subsection{Dependence on structural parameters} \label{sec:anni_param}

\begin{figure}
    \centering
	\includegraphics[]{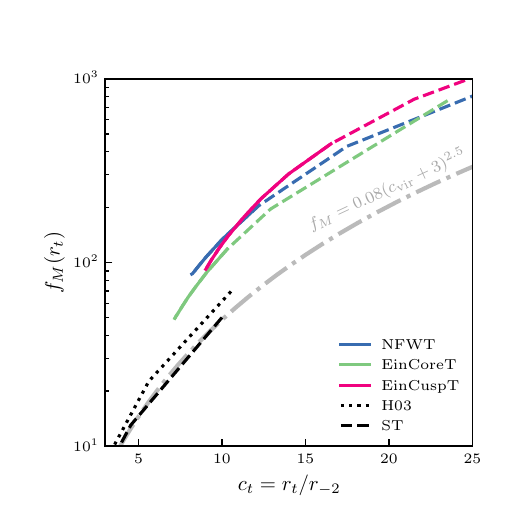}
	\caption{Flux multiplier as a function of the effective concentration parameter, $c_t=r_t/r_{-2}$. For comparison, we show the expected relation for an unstripped NFW profile (dashed-dotted gray line), using the parameterization from \cite{okoli2018}. Note that for the Energy-Truncation models, there are some cases where $r_t>r_{\rm vir}$, and therefore is not a physically accurate depiction of the concentration. We denote cases where $r_t>10 r_s$ with a dashed line. Overall, we find that the Energy Truncation models have flux multipliers  at least twice as large as expected from the unstripped profiles; this is contrary to the expectation that the flux multiplier  of the stripped profile should be approximately equal to that of the unstripped profile within the tidal radius.}
	\label{fig:B_vs_c}
\end{figure}

For an NFW profile, the flux multiplier  can be well-approximated as a function of the concentration parameter $c = r_{\rm vir}/r_{-2}$ alone, with $f \sim c^{2.5}$, while the flux multiplier  for Einasto profiles scales similarly with a slight dependence on the shape parameter $\alpha$ \citep{okoli2018}. However, for tidally stripped haloes, $c_t$ is a more natural concentration parameter, as discussed in Section~\ref{sec:concentration}. In Figure~\ref{fig:B_vs_c}, we show the relationship between the flux multiplier  and the effective concentration parameter, $c_t$. We also compare this relation to the \cite{okoli2018} parameterization for how $f_M$ varies with $c$ for an untruncated NFW profile (the dash-dot grey line). As mentioned in \cite{okoli2018}, we expect that $f_M(c_t)$ should be approximately equal between the unstripped and stripped profiles since most of the mass is removed outside the tidal radius. We find that though $f_M(c_t)$ matches the un-truncated NFW calculation for the H03 and ST models,  our three Energy Truncation models have much higher  flux multipliers for a given tidal radius, as shown previously in Figure~\ref{fig:Boost}. 

The $J$-factor and structural properties of subhaloes are also known to be related. For instance, \cite{delos2019b} examined the relationship between $J$ and $G^{-1} v^4_{\rm max} r^{-1}_{\rm max}$, where $v_{\rm max}$ and  $r_{\rm max}$ are the maximum circular velocity ($v_c = \sqrt{GM(<r)/r}$), and corresponding radius. 
They show that while for an NFW profile $\log_{10} J=1.23 \log_{10} (G^{-1} v^4_{\rm max} r^{-1}_{\rm max})$, their tidally-stripped simulations of NFW profiles have a slightly shallower slope of 0.86. Our NFWT model agrees remarkably well with their findings, but differences in the profile model can cause the $J$-factor to vary by a factor of $\sim2$.

\begin{figure}
    \centering
	\includegraphics[]{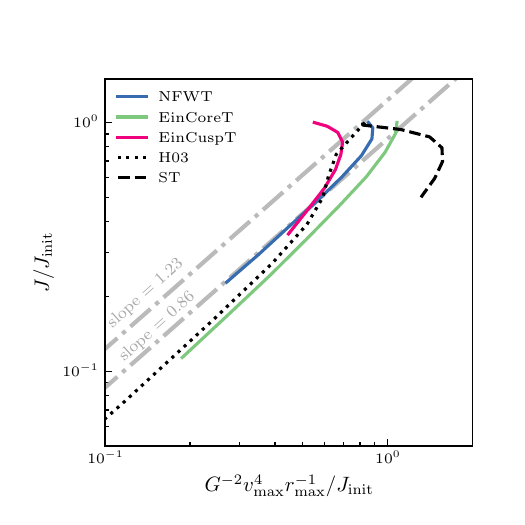}
	\caption{Relationship between the $J$-factor, and the combination of structural parameters $G^{-1} v^4_{\rm max} r^{-1}_{\rm max}$, where $v_{\rm max}$ and  $r_{\rm max}$ are the maximum circular velocity. The slope of 1.23 is what is expected for an NFW profile, while 0.86 is what has been found for tidally stripped NFW profiles \citep{delos2019b}. We find that our NFWT model agrees with this scaling relation, but differences in initial profiles can cause a deviation in the $J$-factor by factors of $\approx 2$.}
	\label{fig:vmax}
\end{figure}

\subsection{Model sensitivity in annihilation: Summary}

In this section, we considered the effects of tidal mass loss on the inner structure of subhaloes as they relate to calculations of the annihilation rate by calculating quantities related to the dark matter annihilation rate for NFW, EinCore, and EinCusp profiles. By using a model assuming a truncation of the distribution function in energy space, we are able to examine a scenario where the central density is conserved during tidal stripping across different initial profile models. We compared our prediction for NFW profiles to two widely used approximations; first, a model in which the original NFW profile is sharply truncated such that it contains the same bound mass (ST model), and secondly, the parameterization from \cite{hayashi2003} that was tuned to simulation results (H03). The ST model will overestimate the density within the tidal radius by construction. By contrast, the Hayashi profile is known to underestimate the central density.

Comparing first the energy-truncated NFW and Einasto profiles, we found that for fixed mass, the concentration, flux multiplier  (measured within the tidal radius), and $J$-factor of the EinCuspT profile is typically 10-50 per cent higher than the NFWT profile, while the EinCoreT profile is typically lower. These differences reflect the different central densities of the initial models, as the annihilation signal is dominated by these high-density regions.  

Then, comparing tidal stripping models (the NFWT model versus the commonly used ST and H03 approximations), we found that neither of the common approximations agrees well with the Energy Truncation model. The effective concentration  at fixed mass is lower by a factor of 10 for both approximations. Both models under-predict the flux multiplier  within the tidal radius by a factor of  $\sim 100$. However, this result is mainly due to the differences in the tidal radii, and if we consider the evolution of $f_M/f_{M, {\rm init}}$ the H03 and ST model underestimate and overestimate the flux multiplier  by $\sim 20$ per cent, respectively. If we consider the $f_M/f_{M, {\rm un}}$; i.e., the flux multiplier  from our model compared to a truncated profile within the same radius, $r_{\rm t}$, we can find that the two approximations underestimate the flux multiplier  by an order of magnitude. Overall, the $J$-factor is expected to be overestimated in the ST model and underestimated in the H03 model; the differences between these models are a factor of 5 once the halo has been stripped to $\sim 5\%$ of its initial mass. 

In summary, subhalo concentrations, flux multipliers , and $J$-factors of tidally stripped haloes are sensitive to the innermost part of the density profile, and predicted fluxes can vary by a factor of 5 or more depending on profile assumptions. Thus, it is important to have physically motivated models that allow us to predict the annihilation of subhaloes down to arbitrarily small bound mass fractions that are not resolved in simulations. The Energy Truncation Model allows us to explore the tidal evolution of the central density of subhaloes under the assumption that cuspy centres are preserved in tidal stripping.
Our results suggest that current constraints on the CDM annihilation cross-section may be inaccurate due to their underlying assumptions about subhalo density profiles; addressing these inaccuracies will be the focus of future work.

\subsection{Implications for the Milky Way} \label{sec:implications}

We provide a brief exploration of what our results mean in the context of the Milky Way galaxy---more robust calculations will be the focus of future work. Throughout this section, we assume that the Milky Way has a virial mass of $M_{\rm MW} = 10^{12}M_{\odot}$, a virial radius of 250 kpc, and a concentration of 10. We use a Planck 2018 cosmology \citep{planck2018}. 

 We assume that the Milky Way is created from the mergers of haloes that follow the distribution ${\rm d}^2 N / {\rm d}M_i {\rm d}z$, where $N$ is the number of haloes, $M_i$ is the mass of a subhalo at infall redshift, $z$. We 
calculate this distribution by assuming the infall population is the same as the field population:
\begin{equation}
    \dfrac{{\rm d}^2N}{{\rm d}M_i {\rm d}z} = \dfrac{1}{\rho_c(z)}  \dfrac{{\rm d} M_{\rm MW}(z)} {{\rm d}z} \dfrac{{\rm d}n}{{\rm d} M } \,\,\, , 
\end{equation}
where ${\rm d}n (M,z)/{\rm d} M $ is the halo mass function from \cite{despali2016},  $M_{\rm MW} (z)$ is the mass of the Milky Way at redshift $z$, and $\rho_c$ is the critical density. Effectively, this specifies the volume of the Universe that must collapse to assemble a halo of mass $M_{\rm MW}$.
Finally, we assume the Milky Way mass evolves according to \cite{wechsler2002}: 
\begin{equation}
M_{\rm MW}(z) = M_{\rm MW} \exp(-8.2z/c) \,\,\, .
\end{equation}

As the Milky Way is formed, some of the substructure will persist as self-bound haloes, with a present day mass of $M_0$. The rest of the dark matter will contribute to the smooth component of the halo. This smooth component can be calculated from:
\begin{equation} \label{eq:M_smooth}
M_{\rm smooth} = \int_\infty^0 \int_{M_{\rm min}}^{M_{\rm MW}} \dfrac{{\rm d}N}{{\rm d}M_i {\rm d}z} (M_i  - M_0) {\rm d}M_i {\rm d}z \,\,\, .
\end{equation}
We assume a minimum halo mass of $10^{-6}\, M_\odot$, as in \citep[][]{okoli2018}. 

We model the relationship between the infall mass and the current mass of the subhaloes using \cite{jiang2014}: 
\begin{equation} \label{eq:Mz}
M_0 =M_i \exp\left( \dfrac{-0.81 \sqrt{200} z_i}{ \pi}\right) \,\,\, .
\end{equation}
We note that this parameterization does not take into account the effect of subhalo mass profile on mass-loss rates, and is likely an overestimate of the tidal mass loss.

Then, the total Milky Way annihilation signal can be expressed as
\begin{equation}
J_{\rm MW} = J_{\rm sub} + J_{\rm smooth} \,\,\, .
\end{equation}
where $J_{\rm sub}$ is the annihilation signal due to Milky Way subhaloes at redshift $z=0$ and $J_{\rm smooth}$ is the annihilation signal from the smooth halo component. 

The signal from the smooth component, $J_{\rm smooth}$, is calculated using the mass from Equation~\eqref{eq:M_smooth}, and assuming an NFW profile. The signal from the substructure can be found by adding the contribution from each subhalo,
\begin{equation}
J_{\rm sub} = \int_\infty^0 \int_{M_{\rm min}}^{M_{\rm MW}} \dfrac{{\rm d}N}{{\rm d}M_i {\rm d}z} J(M_0) {\rm d}M_i {\rm d}z \,\,\, ,
\end{equation}
The annihilation signal of each subhalo, $J = J(M_0)$, depends on the assumed density profile of the subhalo, the subhalo mass, and concentration. We use different models for the subhalo density profiles and assign concentrations to subhaloes, assuming they lie on the median concentration--mass relation from \cite{ishiyama2021}.

\begin{figure} 
    \centering
	\includegraphics[]{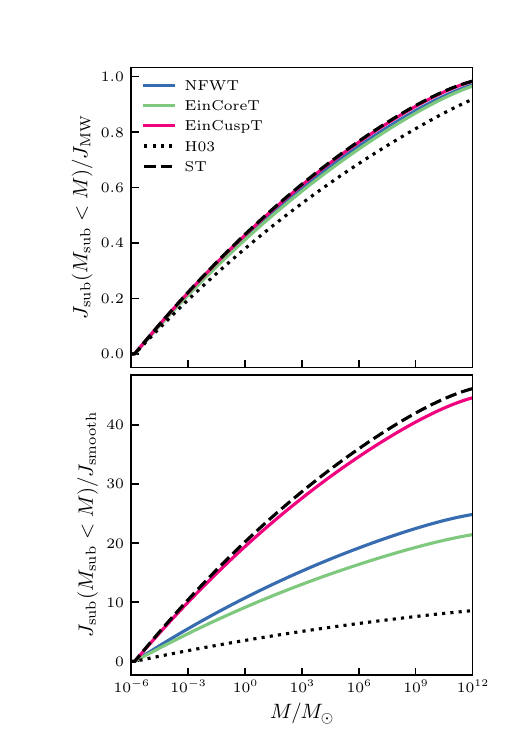}
 \caption{
    \emph{Top:} Cumulative fraction of the Milky Way annihilation signal contributed by substructure, as a function of subhalo mass. The majority of the signal comes from sub-galactic subhaloes, while there is little difference in the mass dependence between the different subhalo models. \emph{Bottom:} `Substructure boost factor', the annihilation rate from the Milky Way including substructure, relative to the rate if all the mass were distributed smoothly as an NFW profile. Including substructure can increase the annihilation rate by a factor of 10--50, depending on the subhalo model used.}
		\label{fig:MW_Ann}
\end{figure}

The top panel of Figure~\ref{fig:MW_Ann} shows the relative contribution of the substructure to the total annihilation signal as a function of subhalo mass. The signal is dominated by the smallest subhaloes, with approximately 70 per cent of the signal coming from haloes less than $10^6 \, M_\odot$. The contribution of subhaloes to the total annihilation signal differs slightly between models,  with small subhaloes in more centrally concentrated models contributing slightly more to the annihilation signal.

Then, we compared the `substructure boost factor' \citep{okoli2018}---defined as the annihilation flux from substructure, $J_{\rm sub}$ relative to the flux from the smooth component of the halo---in the bottom panel Fig~\ref{fig:MW_Ann}. We calculated the smooth component by assuming all of the Milky Way mass was smoothly distributed in an NFW profile with mass\footnote{We find $\log({M_{\rm tot}/M_{\odot}})$ = 11.3 which is slightly lower than our specified $ \log({M_{\rm MW}/M_{\odot}}) $, but agrees to within an order of magnitude.}
\begin{equation}
M_{\rm tot} = \int_\infty^0 \int_{M_{\rm min}}^{M_{\rm max}} \dfrac{{\rm d}N}{{\rm d}M_i {\rm d}z} M_i {\rm d}M_i {\rm d}z \,\,\, .
\end{equation}
{\

We find that including substructure increases the signal by a factor of $\sim50$ for the more centrally concentrated profiles (ST and EinCuspT), and $\sim10$ times for the less centrally concentrated profiles (H03). The lower values are comparable to the predictions of \cite{okoli2018}, who explored different concentration--mass relations, but found a substructure boost factor of $\sim $16 when assuming a H03 model and concentration relations similar to ours\footnote{We note that our quantity is slightly different---we are showing the contribution from substructure relative to a smooth signal calculated using the \emph{total} Milky Way Mass, while \cite{okoli2018} uses the smooth signal from the mass not contained in subhaloes.}. This demonstrates the overall importance of substructure in calculating the annihilation signal in the Milky Way, but also the large range of values predicted depending on the assumed subhalo density profile. Assuming low-mass subhaloes have profiles closest to the EinCuspT or NFWT models, we predict a `substructure boost factor' of as much as 20--50, relative to the contribution from the smooth halo.

We note, however, that there are several simplifications in this analysis. First, our parameterization for subhalo stripping (Equation~\ref{eq:Mz}) overestimates how quickly profiles are disrupted, and does not take into account major mergers. Second, we assumed that all haloes lie on the median concentration--mass relation and that the Milky Way has built up its mass at an average rate. Finally, the rate at which haloes are stripped is dependent on the subhalo model, with cuspier profiles experiencing less stripping---this, however, would \emph{increase} the difference between the models. We leave a full investigation of how  the assumption that cuspy centres are preserved in tidal stripping affects the annihilation signal for future work.

\section{Lensing} \label{sec:lens}

In gravitational lensing, an image of a background galaxy will be distorted by mass along the line of sight.
The mapping between the source and image plane for a background galaxy (i) and its lens (j) is given by the amplification matrix, $a_{ij}$:
\begin{equation}
a_{ij} = \left(\begin{array}{cc} 1-\kappa - \gamma_1 & -\gamma_2\\ -\gamma_2 & 1 -\kappa - \gamma_1\end{array}\right) \,\,\,  ,
\end{equation}
where $\kappa$ is the convergence and $\gamma$ is the shear. In the case of multiple lenses, the total amplification matrix can then be calculated by summing the contributions of each lens.

The radially-averaged convergence and shear profiles are defined as 
\begin{equation}
\begin{gathered}
\kappa(R) = \Sigma(R)/\Sigma_{\rm crit} \,\,\,{\rm and} ,\\
 \gamma(R) = (\bar{\Sigma}(\leq R) - \Sigma(R))/\Sigma_{\rm crit} \,\,\, ,
\end{gathered}
\end{equation} 
and the projected density profile $\Sigma(R)$ can be calculated from the density profile,
\begin{equation}
\Sigma(R) = 2 \int_R^\infty \dfrac{\rho(r) r}{\sqrt{r^2-R^2}}dr \,\,\, ,
\end{equation}
while the mean projected surface mass density is
\begin{equation}
\bar{\Sigma}(\leq R)= \dfrac{M_{\rm proj}(R) } {\pi R^2} \,\,\, ,
\end{equation}
given the projected mass:
\begin{equation}
M_{\rm proj}(R) = 2\pi \int_0^R \Sigma(R') R' dR' \,\,\, ,
\end{equation}
Finally, the critical surface density, $\Sigma_{\rm crit}$ is given by:
\begin{equation}
\Sigma_{\rm crit} = \dfrac{c^2}{4 \pi G} \dfrac{D_S}{D_L D_{LS}} \,\,\, ,
\end{equation}
where $D_S$, $D_L$, and $D_{LS}$ are the angular diameter distances from the observer to the source, the observer to the lens, and the lens to the source, respectively. Since $\Sigma_{\rm crit}$ depends on the geometry of the problem (and consequently the cosmology), it is not well-defined for our isolated simulations. Therefore, in Figure~\ref{fig:Lens} we show $\Sigma$ and $\Delta \Sigma \equiv \bar{\Sigma}-\Sigma$ for our model, as they are proportional to the convergence and shear, respectively; these profiles can then be scaled appropriately given a value of $\Sigma_{\rm crit}$. 
 
 \begin{figure*}
 	\includegraphics[width=1\textwidth]{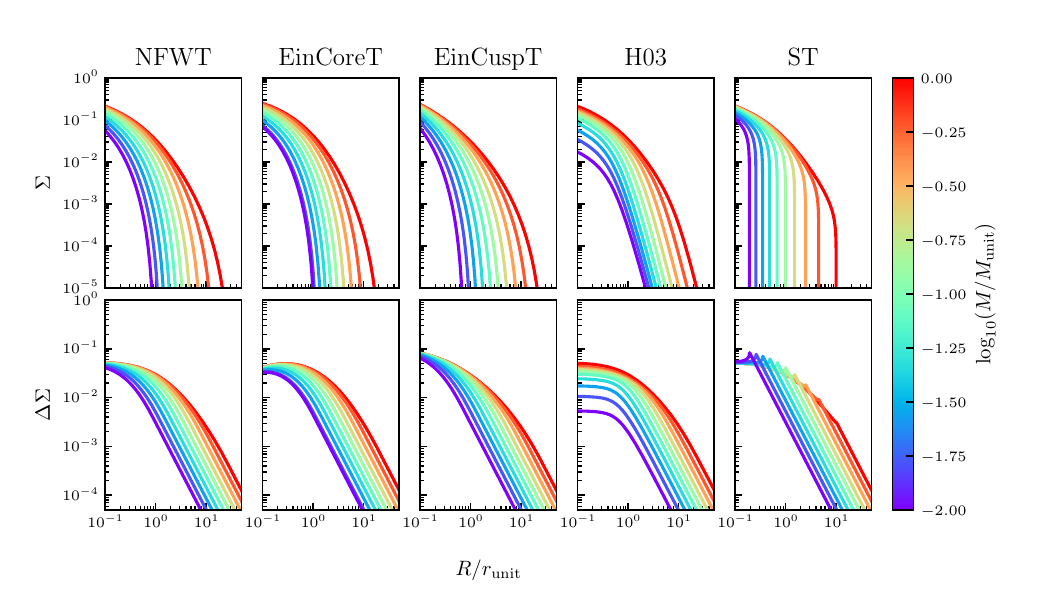}
 	\caption{The projected density profile, $\Sigma$, (top) and $\Delta \Sigma \equiv \bar{\Sigma}-\Sigma$ (bottom) from our models, as a function of bound mass. $\Sigma$ is proportional to the convergence, $\kappa$, and $\Delta \Sigma$ is proportional to the shear $\gamma$ for axisymmetric systems.}
 	\label{fig:Lens}
 \end{figure*}

 In Figure~\ref{fig:Lens}, the projected density profiles (top) show similar trends to the 3D density profiles, with $\Sigma(r>r_t)=0$ (except for H03). The shear profiles (bottom) are approximately constant within the tidal radius and then follow $\Delta \Sigma \propto R^{-2}$. The discontinuity in the slope of the ST model is evident in the shear profiles; the shear is finite but not differentiable at the cutoff radius, as discussed in \cite{baltz2009}.  Figure~\ref{fig:ProjDens} shows the evolution of $\Sigma$ and $\Delta \Sigma$, calculated within $0.5 \, r_{\rm unit}$ and $1\, r_{\rm unit}$, as mass is lost. Initially, the EinCoreT lens models have slightly higher projected densities than the other models, but generally, all models agree to within 10--20 per cent, except at the discontinuity where the tidal radius is close to the aperture size in the ST model. 
    
 \begin{figure*}
 	\includegraphics[width=1\textwidth]{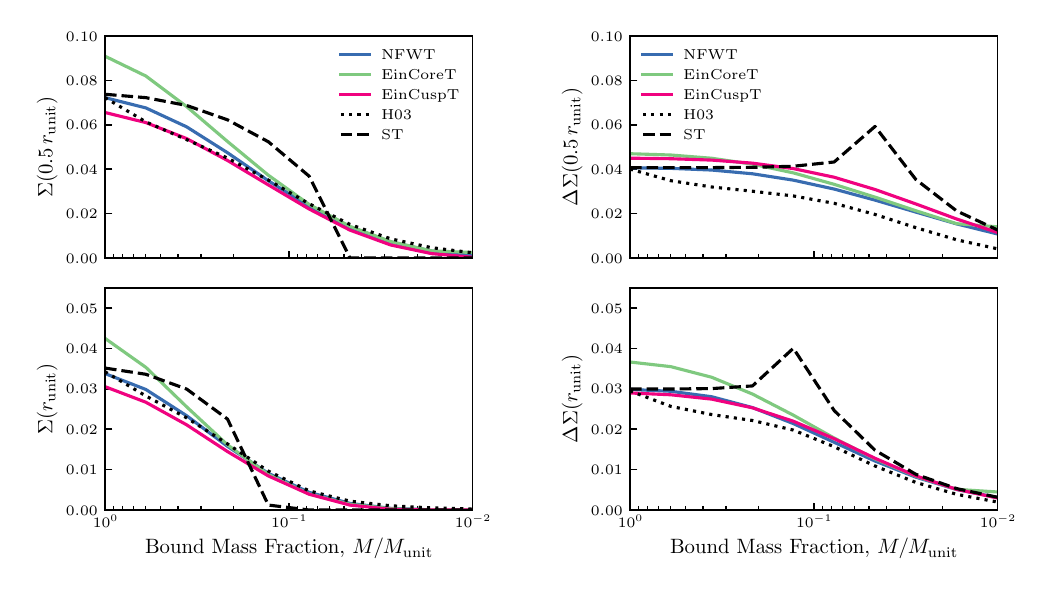}
 	\caption{Variation with mass loss in the mean projected density $\Sigma$ (left) and $\Delta \Sigma$ (right), as measured within apertures of $0.5\,r_{\rm unit}$ (top) or $1\,r_{\rm unit}$ (bottom). Both quantities decrease with mass loss, except for shear in the ST model, which increases when the tidal radius is comparable to the scale radius (initially equal to $1.0\,r_{\rm unit}$). Except at this discontinuity, all models agree to within 10--20 per cent. }
 	\label{fig:ProjDens}
 \end{figure*}

 \subsection{Model comparisons}
 
 To understand better how the convergence and shear depend on the assumed subhalo profile and stripping model, in Figure~\ref{fig:Lens_compare}, we compare radial profiles and residuals for the different models. Between 0.1 $r_{-2}$ and $r_t$,  EinCoreT and EinCuspT mean projected density, $\Sigma$, differ from the NFWT profile by up to 20 per cent, while the shear profiles differ from NFWT profiles by up to 50 per cent. The difference is largest at very small radii, with the EinCuspT profile having higher convergence and sheer values, due to the increased density in the centre of the profile. We also compare the energy-truncated NFW profiles to the commonly used H03 and ST approximations. The H03 model underestimates the convergence and shear profiles at small radii, though differences between H03 and NFWT are comparable to differences found between the different initial profiles, assuming energy truncation. The ST model closely matches the NFWT model at small radii but is less extended. 

  \begin{figure*}
 	\includegraphics[width=1\textwidth]{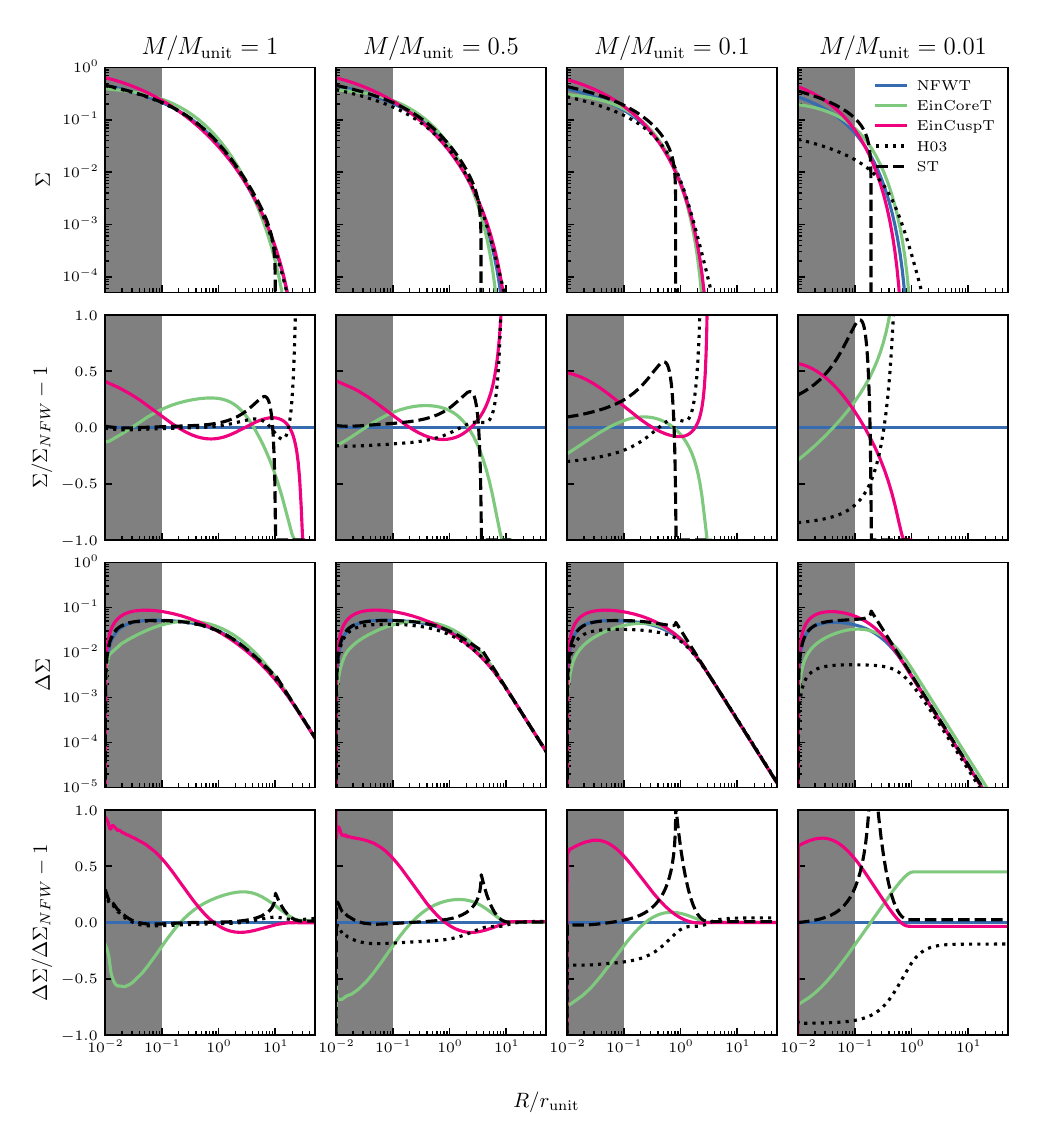}
 	\caption{The projected density profile, $\Sigma$, (top) and $\Delta \Sigma \equiv \bar{\Sigma}-\Sigma$ (third row) 
  for the Energy Truncation model applied to three different profiles (NFWT, EinCoreT and EinCuspT), the stripping model from \cite{hayashi2003} applied to an NFW profile (H03; dotted line), and a sharply truncated NFW profile (ST; dashed line).
   Each column corresponds to a different bound mass, as indicated. The second and last rows show the ratio of $\Sigma$  and $\Delta \Sigma$, respectively, to the NFWT model. The grey box indicates radii that are typically unresolved in isolated simulations. There are significant differences between the three initial halo models, but the ST model does a good job of approximating the NFWT profiles, except for $\Delta\Sigma$ measured close to the truncation radius.}
 	\label{fig:Lens_compare}
 \end{figure*}

Overall, this suggests that errors in the convergence and shear profiles are dominated by assumptions of the initial profile rather than the tidal mass-loss model; i.e., while an ST model does a fairly good job of predicting the convergence and shear of subhalo profiles, different values of the Einasto $\alpha$ parameter can introduce larger differences. Thus, we suspect that using updated tidal models (such as the Energy Truncation model) is unlikely to improve satellite lensing calculations significantly.

Previous work typically finds that different lens models introduce an error of $\sim 10$ per cent in mass estimates \citep[e.g.][]{limousin2005, baltz2009, sereno2016}. 
Additionally, it has been found that even strong tidal truncation has a negligible effect on fitted subhalo parameters  \citep[$M_{200}$ and concentration;][]{minor2021}---however, $M_{200}$ is not necessarily a good estimation of the bound subhalo mass. We explore the effect of profile assumptions on subhalo mass estimates in Figure~\ref{fig:Fit_ProjDens}. We assume that all projected density profiles can be well-described by a sharp truncation of the NFW model (i.e.~the ST model), and fit the free parameters $r_s$, $\rho_0$, and $r_t$. We then calculated the total mass by integrating the projected density profile. We find that the relative error in the subhalo mass typically ranges from 5--10 per cent.

  \begin{figure*}
 	\includegraphics[width=1\textwidth]{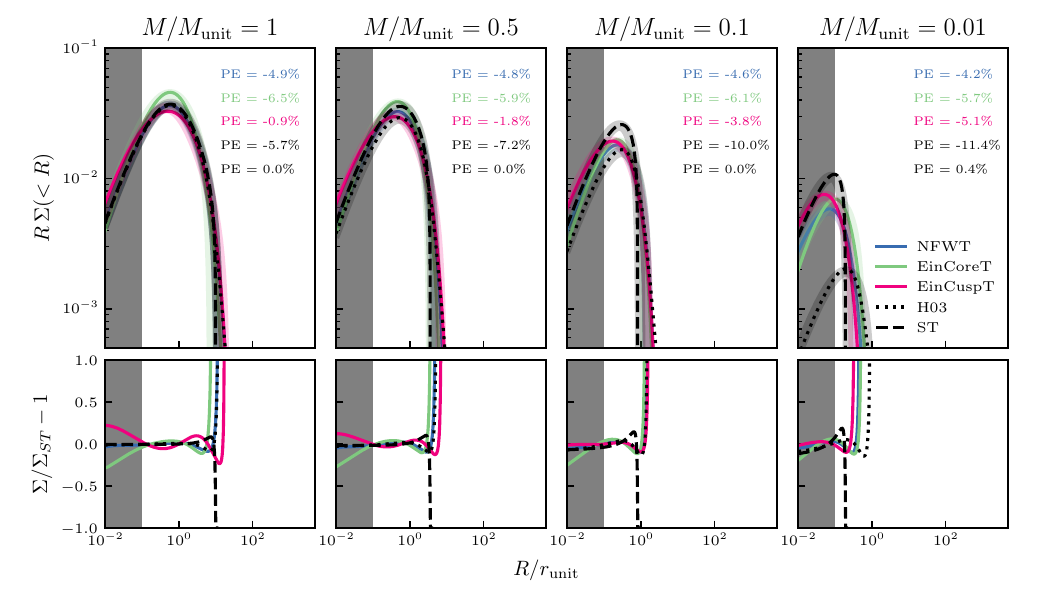}
 \caption{
 \emph{Top:} The projected density profile, $\Sigma$ for the Energy Truncation model applied to three different profiles (NFWT, EinCoreT, and EinCuspT), the stripping model from \cite{hayashi2003} applied to an NFW profile (H03; dotted line), and a sharply truncated NFW profile (ST; dashed line).  Each column corresponds to a different bound mass, as indicated. The thick transparent lines are the best-fit ST models.  We also show the percent error (PE) in the mass estimate calculated as $(M_{\rm ST}-M_{\rm true}) /M_{\rm true} \times 100\%$. \emph{Bottom:} residuals in $\Sigma$ compared to the best-fit ST model.
  }
 	\label{fig:Fit_ProjDens}
 \end{figure*}

 \subsection{Model sensitivity in lensing: Summary}

We find that while there is a significant (30\%\ or more) difference in the projected density and shear between NFW and Einasto profiles with different values of $\alpha$, the tidal stripping model is less important. The central values of lensing observables predicted by the Energy Truncation model are within $\sim$10\%\ of those predicted by the ST (sharp truncation) approximation. In particular, subhalo mass estimates based on ST agree to within 10 per cent, regardless of the mass profile assumed. Thus, we conclude that for lensing applications, modelling results are not sensitive to assumptions about tidal stripping models.

\section{Discussion} \label{sec:discuss}

In this work, we have considered the implications of the uncertainty in subhalo density profiles and tidal evolution for calculations of the dark matter annihilation rate and lensing signals. Specifically, we use the Energy Truncation Model to examine the scenario where cuspy centres are preserved under different subhalo profile assumptions. Overall, we find that the annihilation rate and flux multiplier  are very sensitive to both the subhalo profile and mass loss modelling; at small bound mass ratios, different assumptions lead to annihilation signals that differ by an order of magnitude. While the quantities measured in subhalo lensing also vary slightly, we find that these are generally much less sensitive to assumptions about tidal stripping.

As discussed in \cite{drakos2020,drakos2022}, since subhalo evolution is primarily understood from idealized simulations (due to resolution constraints), predictions for subhalo evolution depend sensitively on the initial halo models assumed in simulations. Here, we have considered Einasto profiles spanning the range of $\alpha$ values one might expect to exist, and also NFW profiles, which are useful for comparison with earlier work. 

Since lower Einasto $\alpha$ values correspond to steeper density profiles, the results in this paper from the EinCusp model are the most relevant for the case of microhaloes, the smallest structures to form in the early universe, whose size is determined by the free-streaming scale of the dark matter particles (although the peak-height dependence of the Einasto shape parameter complicates this statement slightly). 

Assuming dark matter particles have a mass of 100 GeV, the mass of the smallest microhaloes is approximately Earth mass \citep[e.g.][]{diamanti2015}. It is expected that the central density profile of these microhaloes is steeper than that found in larger haloes and thus may contribute greatly to the dark matter annihilation signal \citep[e.g.][]{stucker2023}. We found the concentration of EinCuspT profiles increases rapidly during tidal stripping and has larger annihilation signals than NFWT profiles at a fixed mass. Interestingly, in \cite{drakos2019b}, it was found that Einasto profiles with low $\alpha$ values were the only ones that decreased in concentration after \emph{major} mergers. These results all suggest that profiles with cuspier centres (as found in the earliest haloes) evolve differently.

For instance, haloes may have much higher densities than expected from extrapolations of the low-redshift concentration--mass--redshift relations \cite{ishiyama2014}, and if these densities are conserved, concentrations of the smallest haloes are still uncertain by a factor of $\sim 5$ when extrapolated to low redshift (considering the redshift evolution of the virial radius)\cite{okoli2018}. This uncertainty in concentration translates to an increased flux multiplier  by up to two orders of magnitude. In our work, we find that the Energy Truncation model results in flux multipliers  that are approximately 5 times higher at a given effective concentration (Figure~\ref{fig:B_vs_c}). These results offer separate evidence that the flux multipliers  of the smallest haloes may be greatly underestimated.

While we have focused on dark matter annihilation and lensing applications, subhaloes can also be used to study the nature of dark matter through stellar streams created by disrupted dwarf galaxies. 
For example, stream morphology can help constrain the potential of the host dark matter halo \citep[e.g.][]{dai2018}, and features in streams caused by perturbations from other subhaloes can be used to place constraints on dark matter properties \citep[e.g.][]{carlberg2020}. 
In particular, there has been a lot of interest in studying the GD--1 stellar stream \citep{bonaca2020}, which may have been perturbed by a compact dark matter subhalo. 
Other applications include semi-analytic models of galaxy formation, of which subhalo evolution models are often a key component \cite[e.g.][]{yang2020,jiang2021}. We leave the implications of conserved central density on these applications to future work.

A major assumption in this work is that the Energy Truncation model is an accurate representation of subhalo evolution in the limit that central density is preserved. However, this model has several assumptions that are important to address. First, the infalling subhaloes are assumed to be spherical and isotropic. We expect that as the material is stripped off of subhaloes, the subhalo quickly becomes spherical and isotropic \citep{drakos2020}, and therefore this assumption is approximately valid. However, we plan to explore tidal stripping on non-spherical systems in future work. One way to create non-spherical initial conditions for idealized simulations is by using the output of major merger simulations, as proposed in \citep{moore2004}; in future work, we plan to use the simulations presented in \citep{drakos2019a,drakos2019b} to explore stripping of non-spherical systems. 

Additionally, the concept of "energy" used in the Energy Truncation model is not uniquely defined in a time-dependent, extended system. The binding energy of a particle depends on the chosen frame of reference, the definition of the potential, and the time at which the calculation is performed. While our energy-based ordering is well defined under the assumptions of spherical symmetry and a slowly evolving potential, it is not strictly conserved in more complex or rapidly varying systems. Nonetheless, numerical experiments show that this approach reliably captures the selective stripping of weakly bound particles under gradual mass loss, lending empirical support to its use in the Energy Truncation framework \citep{choi2009, drakos2017, drakos2020}. However, in scenarios involving rapid or impulsive stripping---such as disk shocking or strong subhalo--disk encounters---energy injection can be non-adiabatic, and the assumptions underlying the Energy Truncation model may break down. In such cases, the model may not accurately describe the evolution of subhalo structure, though it may still serve as a useful intermediate baseline between empirical fits like the H03 model and extreme cases such as sharp truncation.

Perhaps the most important assumption made in the Energy Truncation model is that subhaloes only evolve due to tidal fields, and there is no interaction with a baryonic component; extensive work has shown that baryonic matter influences dwarf galaxy properties \citep[as reviewed in][]{bullock2017}. 
It has been proposed that dwarf galaxy haloes can transform from cusps to cores based on the stellar-to-halo mass ratio, $M_*/M_{200}$ \citep{penarrubia2012}, where the cusp is protected from supernova feedback if $M_*/M_{200} \lesssim 5 \times 10^{-4}$ \citep{diCintio2014}, which has been supported by several measurements \citep{read2019,deLeo2023}. These considerations may be relevant for weak lensing studies of substructure since they centre on subhaloes with associated stellar mass by definition. For annihilation studies, on mass scales below the dwarf galaxy range ( $M_{200}<10^{10}\, M_{\odot}$), star formation should be greatly suppressed by photo-heating \citep[e.g.][]{matthias2006, noh2014}. Therefore, it is expected that dark-matter-dominated systems (down to putative Earth-mass-sized haloes) should not be affected by baryonic feedback. These dark matter-dominated systems may contribute greatly to an annihilation signal.

Finally, this work has focused on the effect of subhalo structure on the annihilation signal from individual haloes. To understand what this means, for a complex system like the Milky Way, we would need to model a full population of subhaloes, merging at different times on different orbits and with different concentration parameters. Estimating the signal directly in a self-consistent simulation requires properly correcting for numerical mass loss, which in turn is sensitive to the tidal stripping model assumed. In this paper we have included an initial analysis of the implications for measured dark matter annihilation in the Milky Way. In future work, we plan to track subhalo evolution in a cosmological simulation, using the Energy Truncation model to correct for artificial disruption. We will then use this corrected population to determine the total annihilation rate and place new constraints on proposed dark matter annihilation signals.


\section{Conclusion} \label{sec:conc}
Our best hope for constraining the properties of dark matter through astrophysical observations comes from tests at the highest dark matter densities. Important examples include searches for a gamma-ray or other signal from dark matter annihilation, and tests of substructure in gravitational lenses. The inferences derived from observations depend strongly on the assumed clustering of dark matter on the smallest scales. Unfortunately, this remains unclear from theory; while there has been progress in understanding the average density profile of isolated haloes, it remains unclear how this profile evolves once these become subhaloes. Idealized cases of single mergers have been simulated at increasing resolution, but remain susceptible to numerical effects which can cause increased mass loss \citep[e.g.][]{vandenbosch2018, vandenbosch2018b} and decreased central densities \citep[e.g.][]{errani2020,drakos2022}. To attempt to overcome numerical limitations, we have used a theoretically motivated, universal mass-loss model to study subhalo evolution. We quantify the effect of underestimating the central density of tidally stripped haloes on annihilation signals and lensing profiles of individual subhaloes. 

For lensing mass experiments, we conclude that the details of tidal stripping are less important than assumptions about the original density profile. Thus, recent estimates of the uncertainties in mass introduced by different profiles (approximately 10 per cent) are likely reliable. On the other hand, annihilation constraints are extremely sensitive to uncertainties both in subhalo profiles and in the details of tidal stripping. In particular, our Energy Truncation model predicts annihilation signals that are an order of magnitude larger than some previous estimates, and our initial calculations indicate that the total Milky Way annihilation signal may differ by a factor of five or more depending on the subhalo model used. Future work needs to be done to fully quantify the uncertainties in current dark matter annihilation constraints in light of updated subhalo evolution models. 


\acknowledgments

NED acknowledges support from NSERC Canada, through a postgraduate scholarship. JET acknowledges financial support from NSERC Canada, through a Discovery Grant. \emph{Software}: numpy \citep{numpy}, matplotlib \citep{matplotlib},  scipy \citep{scipy}, \textsc{Colossus}  \citep{colossus}.


\newcommand\aap{A\&A}                
\let\astap=\aap                          
\newcommand\aapr{A\&ARv}             
\newcommand\aaps{A\&AS}              
\newcommand\actaa{Acta Astron.}      
\newcommand\afz{Afz}                 
\newcommand\aj{AJ}                   
\newcommand\ao{Appl. Opt.}           
\let\applopt=\ao                         
\newcommand\aplett{Astrophys.~Lett.} 
\newcommand\apj{ApJ}                 
\newcommand\apjl{ApJ}                
\let\apjlett=\apjl                       
\newcommand\apjs{ApJS}               
\let\apjsupp=\apjs                       
\newcommand\apss{Ap\&SS}             
\newcommand\araa{ARA\&A}             
\newcommand\arep{Astron. Rep.}       
\newcommand\aspc{ASP Conf. Ser.}     
\newcommand\azh{Azh}                 
\newcommand\baas{BAAS}               
\newcommand\bac{Bull. Astron. Inst. Czechoslovakia} 
\newcommand\bain{Bull. Astron. Inst. Netherlands} 
\newcommand\caa{Chinese Astron. Astrophys.} 
\newcommand\cjaa{Chinese J.~Astron. Astrophys.} 
\newcommand\fcp{Fundamentals Cosmic Phys.}  
\newcommand\gca{Geochimica Cosmochimica Acta}   
\newcommand\grl{Geophys. Res. Lett.} 
\newcommand\iaucirc{IAU~Circ.}       
\newcommand\icarus{Icarus}           
\newcommand\japa{J.~Astrophys. Astron.} 
\newcommand\jcap{J.~Cosmology Astropart. Phys.} 
\newcommand\jcp{J.~Chem.~Phys.}      
\newcommand\jgr{J.~Geophys.~Res.}    
\newcommand\jqsrt{J.~Quant. Spectrosc. Radiative Transfer} 
\newcommand\jrasc{J.~R.~Astron. Soc. Canada} 
\newcommand\memras{Mem.~RAS}         
\newcommand\memsai{Mem. Soc. Astron. Italiana} 
\newcommand\mnassa{MNASSA}           
\newcommand\mnras{MNRAS}             
\newcommand\na{New~Astron.}          
\newcommand\nar{New~Astron.~Rev.}    
\newcommand\nat{Nature}              
\newcommand\nphysa{Nuclear Phys.~A}  
\newcommand\pra{Phys. Rev.~A}        
\newcommand\prb{Phys. Rev.~B}        
\newcommand\prc{Phys. Rev.~C}        
\newcommand\prd{Phys. Rev.~D}        
\newcommand\pre{Phys. Rev.~E}        
\newcommand\prl{Phys. Rev.~Lett.}    
\newcommand\pasa{Publ. Astron. Soc. Australia}  
\newcommand\pasp{PASP}               
\newcommand\pasj{PASJ}               
\newcommand\physrep{Phys.~Rep.}      
\newcommand\physscr{Phys.~Scr.}      
\newcommand\planss{Planet. Space~Sci.} 
\newcommand\procspie{Proc.~SPIE}     
\newcommand\rmxaa{Rev. Mex. Astron. Astrofis.} 
\newcommand\qjras{QJRAS}             
\newcommand\sci{Science}             
\newcommand\skytel{Sky \& Telesc.}   
\newcommand\solphys{Sol.~Phys.}      
\newcommand\sovast{Soviet~Ast.}      
\newcommand\ssr{Space Sci. Rev.}     
\newcommand\zap{Z.~Astrophys.}       
\bibliographystyle{JHEP}
\bibliography{TS_app.bib}

\providecommand{\href}[2]{#2}\begingroup\raggedright\begin{thebibliography}{10}

\bibitem{bullock2017}
J.S.~{Bullock} and M.~{Boylan-Kolchin}, \emph{{Small-Scale Challenges to the
  {\ensuremath{\Lambda}}CDM Paradigm}},
  \href{https://doi.org/10.1146/annurev-astro-091916-055313}{\emph{\araa}
  {\bfseries 55} (2017) 343}
  [\href{https://arxiv.org/abs/1707.04256}{{\ttfamily 1707.04256}}].

\bibitem{okoli2018}
C.~{Okoli}, J.E.~{Taylor} and N.~{Afshordi}, \emph{{Searching for dark matter
  annihilation from individual halos: uncertainties, scatter and
  signal-to-noise ratios}},
  \href{https://doi.org/10.1088/1475-7516/2018/08/019}{\emph{\jcap} {\bfseries
  8} (2018) 019} [\href{https://arxiv.org/abs/1711.05271}{{\ttfamily
  1711.05271}}].

\bibitem{ando2019}
S.~{Ando}, T.~{Ishiyama} and N.~{Hiroshima}, \emph{{Halo Substructure Boosts to
  the Signatures of Dark Matter Annihilation}},
  \href{https://doi.org/10.3390/galaxies7030068}{\emph{Galaxies} {\bfseries 7}
  (2019) 68} [\href{https://arxiv.org/abs/1903.11427}{{\ttfamily 1903.11427}}].

\bibitem{taylor2019}
J.E.~{Taylor}, J.~{Shin}, N.N.Q.~{Ouellette} and S.~{Courteau}, \emph{{The
  assembly of the Virgo cluster, traced by its galaxy haloes}},
  \href{https://doi.org/10.1093/mnras/stz1687}{\emph{\mnras} {\bfseries 488}
  (2019) 1111} [\href{https://arxiv.org/abs/1906.07724}{{\ttfamily
  1906.07724}}].

\bibitem{kravtsov2013}
A.V.~{Kravtsov}, \emph{{The Size-Virial Radius Relation of Galaxies}},
  \href{https://doi.org/10.1088/2041-8205/764/2/L31}{\emph{\apjl} {\bfseries
  764} (2013) L31} [\href{https://arxiv.org/abs/1212.2980}{{\ttfamily
  1212.2980}}].

\bibitem{somerville2018}
R.S.~{Somerville}, P.~{Behroozi}, V.~{Pandya}, A.~{Dekel}, S.M.~{Faber},
  A.~{Fontana} et~al., \emph{{The relationship between galaxy and dark matter
  halo size from z {\ensuremath{\sim}} 3 to the present}},
  \href{https://doi.org/10.1093/mnras/stx2040}{\emph{\mnras} {\bfseries 473}
  (2018) 2714} [\href{https://arxiv.org/abs/1701.03526}{{\ttfamily
  1701.03526}}].

\bibitem{chang2021}
L.J.~{Chang} and L.~{Necib}, \emph{{Dark matter density profiles in dwarf
  galaxies: linking Jeans modelling systematics and observation}},
  \href{https://doi.org/10.1093/mnras/stab2440}{\emph{\mnras} {\bfseries 507}
  (2021) 4715} [\href{https://arxiv.org/abs/2009.00613}{{\ttfamily
  2009.00613}}].

\bibitem{natarajan2002}
P.~{Natarajan}, J.-P.~{Kneib} and I.~{Smail}, \emph{{Evidence for Tidal
  Stripping of Dark Matter Halos in Massive Cluster Lenses}},
  \href{https://doi.org/10.1086/345399}{\emph{\apjl} {\bfseries 580} (2002)
  L11} [\href{https://arxiv.org/abs/astro-ph/0207049}{{\ttfamily
  astro-ph/0207049}}].

\bibitem{li2016}
R.~{Li}, H.~{Shan}, J.-P.~{Kneib}, H.~{Mo}, E.~{Rozo}, A.~{Leauthaud} et~al.,
  \emph{{Measuring subhalo mass in redMaPPer clusters with CFHT Stripe 82
  Survey}}, \href{https://doi.org/10.1093/mnras/stw494}{\emph{\mnras}
  {\bfseries 458} (2016) 2573}
  [\href{https://arxiv.org/abs/1507.01464}{{\ttfamily 1507.01464}}].

\bibitem{niemiec2017}
A.~{Niemiec}, E.~{Jullo}, M.~{Limousin}, C.~{Giocoli}, T.~{Erben},
  H.~{Hildebrant} et~al., \emph{{Stellar-to-halo mass relation of cluster
  galaxies}}, \href{https://doi.org/10.1093/mnras/stx1667}{\emph{\mnras}
  {\bfseries 471} (2017) 1153}
  [\href{https://arxiv.org/abs/1703.03348}{{\ttfamily 1703.03348}}].

\bibitem{sifon2018}
C.~{Sif{\'o}n}, R.~{Herbonnet}, H.~{Hoekstra}, R.F.J.~{van der Burg} and
  M.~{Viola}, \emph{{The galaxy-subhalo connection in low-redshift galaxy
  clusters from weak gravitational lensing}},
  \href{https://doi.org/10.1093/mnras/sty1161}{\emph{\mnras} {\bfseries 478}
  (2018) 1244} [\href{https://arxiv.org/abs/1706.06125}{{\ttfamily
  1706.06125}}].

\bibitem{dvornik2020}
A.~{Dvornik}, H.~{Hoekstra}, K.~{Kuijken}, A.H.~{Wright}, M.~{Asgari},
  M.~{Bilicki} et~al., \emph{{KiDS+GAMA: The weak lensing calibrated
  stellar-to-halo mass relation of central and satellite galaxies}},
  \href{https://doi.org/10.1051/0004-6361/202038693}{\emph{\aap} {\bfseries
  642} (2020) A83} [\href{https://arxiv.org/abs/2006.10777}{{\ttfamily
  2006.10777}}].

\bibitem{kumar2022}
A.~{Kumar}, S.~{More} and D.~{Rana}, \emph{{Subaru HSC weak lensing of SDSS
  redMaPPer cluster satellite galaxies: empirical upper limit on orphan
  fractions}}, \href{https://doi.org/10.1093/mnras/stac2862}{\emph{\mnras}
  {\bfseries 517} (2022) 4389}
  [\href{https://arxiv.org/abs/2205.00018}{{\ttfamily 2205.00018}}].

\bibitem{wang2024}
C.~{Wang}, R.~{Li}, H.~{Shan}, W.~{Xu}, J.~{Yao}, Y.~{Jing} et~al.,
  \emph{{Assessing mass-loss and stellar-to-halo mass ratio of satellite
  galaxies: a galaxy-galaxy lensing approach utilizing DECaLS DR8 data}},
  \href{https://doi.org/10.1093/mnras/stae121}{\emph{\mnras} {\bfseries 528}
  (2024) 2728} [\href{https://arxiv.org/abs/2305.13694}{{\ttfamily
  2305.13694}}].

\bibitem{baltz2009}
E.A.~{Baltz}, P.~{Marshall} and M.~{Oguri}, \emph{{Analytic models of plausible
  gravitational lens potentials}},
  \href{https://doi.org/10.1088/1475-7516/2009/01/015}{\emph{\jcap} {\bfseries
  2009} (2009) 015} [\href{https://arxiv.org/abs/0705.0682}{{\ttfamily
  0705.0682}}].

\bibitem{hayashi2003}
E.~{Hayashi}, J.F.~{Navarro}, J.E.~{Taylor}, J.~{Stadel} and T.~{Quinn},
  \emph{{The Structural Evolution of Substructure}},
  \href{https://doi.org/10.1086/345788}{\emph{\apj} {\bfseries 584} (2003) 541}
  [\href{https://arxiv.org/abs/astro-ph/0203004}{{\ttfamily
  astro-ph/0203004}}].

\bibitem{diemand2007}
J.~{Diemand}, M.~{Kuhlen} and P.~{Madau}, \emph{{Dark Matter Substructure and
  Gamma-Ray Annihilation in the Milky Way Halo}},
  \href{https://doi.org/10.1086/510736}{\emph{\apj} {\bfseries 657} (2007) 262}
  [\href{https://arxiv.org/abs/astro-ph/0611370}{{\ttfamily
  astro-ph/0611370}}].

\bibitem{vandenbosch2017}
F.C.~{van den Bosch}, \emph{{Dissecting the evolution of dark matter subhaloes
  in the Bolshoi simulation}},
  \href{https://doi.org/10.1093/mnras/stx520}{\emph{\mnras} {\bfseries 468}
  (2017) 885} [\href{https://arxiv.org/abs/1611.02657}{{\ttfamily
  1611.02657}}].

\bibitem{vandenbosch2018}
F.C.~{van den Bosch}, G.~{Ogiya}, O.~{Hahn} and A.~{Burkert}, \emph{{Disruption
  of dark matter substructure: fact or fiction?}},
  \href{https://doi.org/10.1093/mnras/stx2956}{\emph{\mnras} {\bfseries 474}
  (2018) 3043} [\href{https://arxiv.org/abs/1711.05276}{{\ttfamily
  1711.05276}}].

\bibitem{vandenbosch2018b}
F.C.~{van den Bosch} and G.~{Ogiya}, \emph{{Dark matter substructure in
  numerical simulations: a tale of discreteness noise, runaway instabilities,
  and artificial disruption}},
  \href{https://doi.org/10.1093/mnras/sty084}{\emph{\mnras} {\bfseries 475}
  (2018) 4066} [\href{https://arxiv.org/abs/1801.05427}{{\ttfamily
  1801.05427}}].

\bibitem{benson2022}
A.J.~{Benson} and X.~{Du}, \emph{{Tidal tracks and artificial disruption of
  cold dark matter haloes}},
  \href{https://doi.org/10.1093/mnras/stac2750}{\emph{\mnras} {\bfseries 517}
  (2022) 1398} [\href{https://arxiv.org/abs/2206.01842}{{\ttfamily
  2206.01842}}].

\bibitem{kazantzidis2006}
S.~{Kazantzidis}, A.R.~{Zentner} and A.V.~{Kravtsov}, \emph{{The Robustness of
  Dark Matter Density Profiles in Dissipationless Mergers}},
  \href{https://doi.org/10.1086/500579}{\emph{\apj} {\bfseries 641} (2006) 647}
  [\href{https://arxiv.org/abs/astro-ph/0510583}{{\ttfamily
  astro-ph/0510583}}].

\bibitem{errani2020}
R.~{Errani} and J.~{Pe{\~n}arrubia}, \emph{{Can tides disrupt cold dark matter
  subhaloes?}}, \href{https://doi.org/10.1093/mnras/stz3349}{\emph{\mnras}
  {\bfseries 491} (2020) 4591}
  [\href{https://arxiv.org/abs/1906.01642}{{\ttfamily 1906.01642}}].

\bibitem{amorisco2021}
N.C.~{Amorisco}, \emph{{Cold dark matter subhaloes at arbitrarily low masses}},
  \href{https://doi.org/10.48550/arXiv.2111.01148}{\emph{arXiv e-prints} (2021)
  arXiv:2111.01148} [\href{https://arxiv.org/abs/2111.01148}{{\ttfamily
  2111.01148}}].

\bibitem{drakos2022}
N.E.~{Drakos}, J.E.~{Taylor} and A.J.~{Benson}, \emph{{A universal model for
  the evolution of tidally stripped systems}},
  \href{https://doi.org/10.1093/mnras/stac2202}{\emph{\mnras} {\bfseries 516}
  (2022) 106} [\href{https://arxiv.org/abs/2207.14803}{{\ttfamily
  2207.14803}}].

\bibitem{ishiyama2014}
T.~{Ishiyama}, \emph{{Hierarchical Formation of Dark Matter Halos and the Free
  Streaming Scale}},
  \href{https://doi.org/10.1088/0004-637X/788/1/27}{\emph{\apj} {\bfseries 788}
  (2014) 27} [\href{https://arxiv.org/abs/1404.1650}{{\ttfamily 1404.1650}}].

\bibitem{angulo2017}
R.E.~{Angulo}, O.~{Hahn}, A.D.~{Ludlow} and S.~{Bonoli}, \emph{{Earth-mass
  haloes and the emergence of NFW density profiles}},
  \href{https://doi.org/10.1093/mnras/stx1658}{\emph{\mnras} {\bfseries 471}
  (2017) 4687} [\href{https://arxiv.org/abs/1604.03131}{{\ttfamily
  1604.03131}}].

\bibitem{ogiya2018}
G.~{Ogiya} and O.~{Hahn}, \emph{{What sets the central structure of dark matter
  haloes?}}, \href{https://doi.org/10.1093/mnras/stx2639}{\emph{\mnras}
  {\bfseries 473} (2018) 4339}
  [\href{https://arxiv.org/abs/1707.07693}{{\ttfamily 1707.07693}}].

\bibitem{delos2019a}
M.S.~{Delos}, M.~{Bruff} and A.L.~{Erickcek}, \emph{{Predicting the density
  profiles of the first halos}},
  \href{https://doi.org/10.1103/PhysRevD.100.023523}{\emph{\prd} {\bfseries
  100} (2019) 023523} [\href{https://arxiv.org/abs/1905.05766}{{\ttfamily
  1905.05766}}].

\bibitem{delos2023}
M.S.~{Delos} and S.D.M.~{White}, \emph{{Inner cusps of the first dark matter
  haloes: formation and survival in a cosmological context}},
  \href{https://doi.org/10.1093/mnras/stac3373}{\emph{\mnras} {\bfseries 518}
  (2023) 3509} [\href{https://arxiv.org/abs/2207.05082}{{\ttfamily
  2207.05082}}].

\bibitem{delos2023b}
M.S.~{Delos}, M.~{Korsmeier}, A.~{Widmark}, C.~{Blanco}, T.~{Linden} and
  S.D.M.~{White}, \emph{{Limits on dark matter annihilation in prompt cusps
  from the isotropic gamma-ray background}},
  \href{https://doi.org/10.1103/PhysRevD.109.083512}{\emph{\prd} {\bfseries
  109} (2024) 083512} [\href{https://arxiv.org/abs/2307.13023}{{\ttfamily
  2307.13023}}].

\bibitem{delos2022}
M.S.~{Delos} and S.D.M.~{White}, \emph{{Prompt cusps and the dark matter
  annihilation signal}},
  \href{https://doi.org/10.1088/1475-7516/2023/10/008}{\emph{\jcap} {\bfseries
  2023} (2023) 008} [\href{https://arxiv.org/abs/2209.11237}{{\ttfamily
  2209.11237}}].

\bibitem{stucker2023}
J.~{St{\"u}cker}, G.~{Ogiya}, S.D.M.~{White} and R.E.~{Angulo}, \emph{{The
  effect of stellar encounters on the dark matter annihilation signal from
  prompt cusps}}, \href{https://doi.org/10.1093/mnras/stad1268}{\emph{\mnras}
  {\bfseries 523} (2023) 1067}
  [\href{https://arxiv.org/abs/2301.04670}{{\ttfamily 2301.04670}}].

\bibitem{navarro1996}
J.F.~{Navarro}, C.S.~{Frenk} and S.D.M.~{White}, \emph{{The Structure of Cold
  Dark Matter Halos}}, \href{https://doi.org/10.1086/177173}{\emph{\apj}
  {\bfseries 462} (1996) 563}
  [\href{https://arxiv.org/abs/astro-ph/9508025}{{\ttfamily
  astro-ph/9508025}}].

\bibitem{navarro1997}
J.F.~{Navarro}, C.S.~{Frenk} and S.D.M.~{White}, \emph{{A Universal Density
  Profile from Hierarchical Clustering}}, {\emph{\apj} {\bfseries 490} (1997)
  493} [\href{https://arxiv.org/abs/astro-ph/9611107}{{\ttfamily
  astro-ph/9611107}}].

\bibitem{einasto1965}
J.~{Einasto}, \emph{{Kinematics and dynamics of stellar systems}}, {\emph{Trudy
  Inst. Astrofiz. Alma-Ata} {\bfseries 5} (1965) 87}.

\bibitem{navarro2004}
J.F.~{Navarro}, E.~{Hayashi}, C.~{Power}, A.R.~{Jenkins}, C.S.~{Frenk},
  S.D.M.~{White} et~al., \emph{{The inner structure of {$\Lambda$}CDM haloes -
  III. Universality and asymptotic slopes}},
  \href{https://doi.org/10.1111/j.1365-2966.2004.07586.x}{\emph{\mnras}
  {\bfseries 349} (2004) 1039}
  [\href{https://arxiv.org/abs/astro-ph/0311231}{{\ttfamily
  astro-ph/0311231}}].

\bibitem{gao2008}
L.~{Gao}, J.F.~{Navarro}, S.~{Cole}, C.S.~{Frenk}, S.D.M.~{White},
  V.~{Springel} et~al., \emph{{The redshift dependence of the structure of
  massive {$\Lambda$} cold dark matter haloes}},
  \href{https://doi.org/10.1111/j.1365-2966.2008.13277.x}{\emph{\mnras}
  {\bfseries 387} (2008) 536}
  [\href{https://arxiv.org/abs/0711.0746}{{\ttfamily 0711.0746}}].

\bibitem{klypin2016}
A.~{Klypin}, G.~{Yepes}, S.~{Gottl{\"o}ber}, F.~{Prada} and S.~{He{\ss}},
  \emph{{MultiDark simulations: the story of dark matter halo concentrations
  and density profiles}},
  \href{https://doi.org/10.1093/mnras/stw248}{\emph{\mnras} {\bfseries 457}
  (2016) 4340} [\href{https://arxiv.org/abs/1411.4001}{{\ttfamily 1411.4001}}].

\bibitem{kazantzidis2004}
S.~{Kazantzidis}, J.~{Magorrian} and B.~{Moore}, \emph{{Generating Equilibrium
  Dark Matter Halos: Inadequacies of the Local Maxwellian Approximation}},
  \href{https://doi.org/10.1086/380192}{\emph{\apj} {\bfseries 601} (2004) 37}
  [\href{https://arxiv.org/abs/astro-ph/0309517}{{\ttfamily
  astro-ph/0309517}}].

\bibitem{penarrubia2010}
J.~{Pe{\~n}arrubia}, A.J.~{Benson}, M.G.~{Walker}, G.~{Gilmore},
  A.W.~{McConnachie} and L.~{Mayer}, \emph{{The impact of dark matter cusps and
  cores on the satellite galaxy population around spiral galaxies}},
  \href{https://doi.org/10.1111/j.1365-2966.2010.16762.x}{\emph{\mnras}
  {\bfseries 406} (2010) 1290}
  [\href{https://arxiv.org/abs/1002.3376}{{\ttfamily 1002.3376}}].

\bibitem{green2019}
S.B.~{Green} and F.C.~{van den Bosch}, \emph{{The tidal evolution of dark
  matter substructure - I. subhalo density profiles}},
  \href{https://doi.org/10.1093/mnras/stz2767}{\emph{\mnras} {\bfseries 490}
  (2019) 2091} [\href{https://arxiv.org/abs/1908.08537}{{\ttfamily
  1908.08537}}].

\bibitem{drakos2017}
N.E.~{Drakos}, J.E.~{Taylor} and A.J.~{Benson}, \emph{{The phase-space
  structure of tidally stripped haloes}},
  \href{https://doi.org/10.1093/mnras/stx652}{\emph{\mnras} {\bfseries 468}
  (2017) 2345} [\href{https://arxiv.org/abs/1703.07836}{{\ttfamily
  1703.07836}}].

\bibitem{drakos2020}
N.E.~{Drakos}, J.E.~{Taylor} and A.J.~{Benson}, \emph{{Mass loss in tidally
  stripped systems; the energy-based truncation method}},
  \href{https://doi.org/10.1093/mnras/staa760}{\emph{\mnras} {\bfseries 494}
  (2020) 378} [\href{https://arxiv.org/abs/2003.09452}{{\ttfamily
  2003.09452}}].

\bibitem{widrow2005}
L.M.~{Widrow} and J.~{Dubinski}, \emph{{Equilibrium Disk-Bulge-Halo Models for
  the Milky Way and Andromeda Galaxies}},
  \href{https://doi.org/10.1086/432710}{\emph{\apj} {\bfseries 631} (2005) 838}
  [\href{https://arxiv.org/abs/astro-ph/0506177}{{\ttfamily
  astro-ph/0506177}}].

\bibitem{choi2009}
J.-H.~{Choi}, M.D.~{Weinberg} and N.~{Katz}, \emph{{The dynamics of satellite
  disruption in cold dark matter haloes}},
  \href{https://doi.org/10.1111/j.1365-2966.2009.15556.x}{\emph{\mnras}
  {\bfseries 400} (2009) 1247}
  [\href{https://arxiv.org/abs/0812.0009}{{\ttfamily 0812.0009}}].

\bibitem{stucker2021}
J.~{St{\"u}cker}, R.E.~{Angulo} and P.~{Busch}, \emph{{The boosted potential}},
  \href{https://doi.org/10.1093/mnras/stab2913}{\emph{\mnras} {\bfseries 508}
  (2021) 5196} [\href{https://arxiv.org/abs/2107.13008}{{\ttfamily
  2107.13008}}].

\bibitem{eddington1916}
A.S.~{Eddington}, \emph{{The distribution of stars in globular clusters}},
  \href{https://doi.org/10.1093/mnras/76.7.572}{\emph{\mnras} {\bfseries 76}
  (1916) 572}.

\bibitem{king1966}
I.R.~{King}, \emph{{The structure of star clusters. III. Some simple dynamical
  models}}, \href{https://doi.org/10.1086/109857}{\emph{\aj} {\bfseries 71}
  (1966) 64}.

\bibitem{lazar2023}
A.~{Lazar}, J.S.~{Bullock}, A.~{Nierenberg}, L.A.~{Moustakas} and
  M.~{Boylan-Kolchin}, \emph{{An analytic surface density profile for
  {\ensuremath{\Lambda}}CDM haloes and gravitational lensing studies}},
  \href{https://doi.org/10.1093/mnras/stae035}{\emph{\mnras} {\bfseries 528}
  (2024) 444} [\href{https://arxiv.org/abs/2304.11177}{{\ttfamily
  2304.11177}}].

\bibitem{drakos2019b}
N.E.~{Drakos}, J.E.~{Taylor}, A.~{Berrouet}, A.S.G.~{Robotham} and C.~{Power},
  \emph{{Major mergers between dark matter haloes - II. Profile and
  concentration changes}},
  \href{https://doi.org/10.1093/mnras/stz1307}{\emph{\mnras} {\bfseries 487}
  (2019) 1008} [\href{https://arxiv.org/abs/1811.12844}{{\ttfamily
  1811.12844}}].

\bibitem{wechsler2002}
R.H.~{Wechsler}, J.S.~{Bullock}, J.R.~{Primack}, A.V.~{Kravtsov} and
  A.~{Dekel}, \emph{{Concentrations of Dark Halos from Their Assembly
  Histories}}, \href{https://doi.org/10.1086/338765}{\emph{\apj} {\bfseries
  568} (2002) 52} [\href{https://arxiv.org/abs/astro-ph/0108151}{{\ttfamily
  astro-ph/0108151}}].

\bibitem{zhao2003}
D.H.~{Zhao}, H.J.~{Mo}, Y.P.~{Jing} and G.~{B{\"o}rner}, \emph{{The growth and
  structure of dark matter haloes}},
  \href{https://doi.org/10.1046/j.1365-8711.2003.06135.x}{\emph{\mnras}
  {\bfseries 339} (2003) 12}
  [\href{https://arxiv.org/abs/astro-ph/0204108}{{\ttfamily
  astro-ph/0204108}}].

\bibitem{zhao2009}
D.H.~{Zhao}, Y.P.~{Jing}, H.J.~{Mo} and G.~{B{\"o}rner}, \emph{{Accurate
  Universal Models for the Mass Accretion Histories and Concentrations of Dark
  Matter Halos}},
  \href{https://doi.org/10.1088/0004-637X/707/1/354}{\emph{\apj} {\bfseries
  707} (2009) 354} [\href{https://arxiv.org/abs/0811.0828}{{\ttfamily
  0811.0828}}].

\bibitem{wong2012}
A.W.C.~{Wong} and J.E.~{Taylor}, \emph{{What Do Dark Matter Halo Properties
  Tell Us about Their Mass Assembly Histories?}},
  \href{https://doi.org/10.1088/0004-637X/757/1/102}{\emph{\apj} {\bfseries
  757} (2012) 102} [\href{https://arxiv.org/abs/1112.4229}{{\ttfamily
  1112.4229}}].

\bibitem{ludlow2014}
A.D.~{Ludlow}, J.F.~{Navarro}, R.E.~{Angulo}, M.~{Boylan-Kolchin},
  V.~{Springel}, C.~{Frenk} et~al., \emph{{The mass-concentration-redshift
  relation of cold dark matter haloes}},
  \href{https://doi.org/10.1093/mnras/stu483}{\emph{\mnras} {\bfseries 441}
  (2014) 378} [\href{https://arxiv.org/abs/1312.0945}{{\ttfamily 1312.0945}}].

\bibitem{correa2015}
C.A.~{Correa}, J.S.B.~{Wyithe}, J.~{Schaye} and A.R.~{Duffy}, \emph{{The
  accretion history of dark matter haloes - III. A physical model for the
  concentration-mass relation}},
  \href{https://doi.org/10.1093/mnras/stv1363}{\emph{\mnras} {\bfseries 452}
  (2015) 1217} [\href{https://arxiv.org/abs/1502.00391}{{\ttfamily
  1502.00391}}].

\bibitem{wang2020}
K.~{Wang}, Y.-Y.~{Mao}, A.R.~{Zentner}, J.U.~{Lange}, F.C.~{van den Bosch} and
  R.H.~{Wechsler}, \emph{{Concentrations of dark haloes emerge from their
  merger histories}},
  \href{https://doi.org/10.1093/mnras/staa2733}{\emph{\mnras} {\bfseries 498}
  (2020) 4450} [\href{https://arxiv.org/abs/2004.13732}{{\ttfamily
  2004.13732}}].

\bibitem{bartels2015}
R.~{Bartels} and S.~{Ando}, \emph{{Boosting the annihilation boost: Tidal
  effects on dark matter subhalos and consistent luminosity modeling}},
  \href{https://doi.org/10.1103/PhysRevD.92.123508}{\emph{\prd} {\bfseries 92}
  (2015) 123508} [\href{https://arxiv.org/abs/1507.08656}{{\ttfamily
  1507.08656}}].

\bibitem{han2016}
J.~{Han}, S.~{Cole}, C.S.~{Frenk} and Y.~{Jing}, \emph{{A unified model for the
  spatial and mass distribution of subhaloes}},
  \href{https://doi.org/10.1093/mnras/stv2900}{\emph{\mnras} {\bfseries 457}
  (2016) 1208} [\href{https://arxiv.org/abs/1509.02175}{{\ttfamily
  1509.02175}}].

\bibitem{stref2017}
M.~{Stref} and J.~{Lavalle}, \emph{{Modeling dark matter subhalos in a
  constrained galaxy: Global mass and boosted annihilation profiles}},
  \href{https://doi.org/10.1103/PhysRevD.95.063003}{\emph{\prd} {\bfseries 95}
  (2017) 063003} [\href{https://arxiv.org/abs/1610.02233}{{\ttfamily
  1610.02233}}].

\bibitem{hiroshima2018}
N.~{Hiroshima}, S.~{Ando} and T.~{Ishiyama}, \emph{{Modeling evolution of dark
  matter substructure and annihilation boost}},
  \href{https://doi.org/10.1103/PhysRevD.97.123002}{\emph{\prd} {\bfseries 97}
  (2018) 123002} [\href{https://arxiv.org/abs/1803.07691}{{\ttfamily
  1803.07691}}].

\bibitem{hutten2019}
M.~{H{\"u}tten}, M.~{Stref}, C.~{Combet}, J.~{Lavalle} and D.~{Maurin},
  \emph{{{\ensuremath{\gamma}}-ray and {\ensuremath{\nu}} Searches for
  Dark-Matter Subhalos in the Milky Way with a Baryonic Potential}},
  \href{https://doi.org/10.3390/galaxies7020060}{\emph{Galaxies} {\bfseries 7}
  (2019) 60} [\href{https://arxiv.org/abs/1904.10935}{{\ttfamily 1904.10935}}].

\bibitem{ibarra2019}
A.~{Ibarra}, B.J.~{Kavanagh} and A.~{Rappelt}, \emph{{Impact of substructure on
  local dark matter searches}},
  \href{https://doi.org/10.1088/1475-7516/2019/12/013}{\emph{\jcap} {\bfseries
  2019} (2019) 013} [\href{https://arxiv.org/abs/1908.00747}{{\ttfamily
  1908.00747}}].

\bibitem{delos2019d}
M.S.~{Delos}, T.~{Linden} and A.L.~{Erickcek}, \emph{{Breaking a dark
  degeneracy: The gamma-ray signature of early matter domination}},
  \href{https://doi.org/10.1103/PhysRevD.100.123546}{\emph{\prd} {\bfseries
  100} (2019) 123546} [\href{https://arxiv.org/abs/1910.08553}{{\ttfamily
  1910.08553}}].

\bibitem{stref2019}
M.~{Stref}, T.~{Lacroix} and J.~{Lavalle}, \emph{{Remnants of Galactic Subhalos
  and Their Impact on Indirect Dark-Matter Searches}},
  \href{https://doi.org/10.3390/galaxies7020065}{\emph{Galaxies} {\bfseries 7}
  (2019) 65} [\href{https://arxiv.org/abs/1905.02008}{{\ttfamily 1905.02008}}].

\bibitem{facchinetti2022}
G.~{Facchinetti}, J.~{Lavalle} and M.~{Stref}, \emph{{Statistics for dark
  matter subhalo searches in gamma rays from a kinematically constrained
  population model: Fermi-LAT-like telescopes}},
  \href{https://doi.org/10.1103/PhysRevD.106.083023}{\emph{\prd} {\bfseries
  106} (2022) 083023}.

\bibitem{taylor2003}
J.E.~{Taylor} and J.~{Silk}, \emph{{The clumpiness of cold dark matter:
  implications for the annihilation signal}},
  \href{https://doi.org/10.1046/j.1365-8711.2003.06201.x}{\emph{\mnras}
  {\bfseries 339} (2003) 505}
  [\href{https://arxiv.org/abs/astro-ph/0207299}{{\ttfamily
  astro-ph/0207299}}].

\bibitem{bergstrom1998}
L.~{Bergstr{\"o}m}, P.~{Ullio} and J.H.~{Buckley}, \emph{{Observability of
  {\ensuremath{\gamma}} rays from dark matter neutralino annihilations in the
  Milky Way halo}},
  \href{https://doi.org/10.1016/S0927-6505(98)00015-2}{\emph{Astroparticle
  Physics} {\bfseries 9} (1998) 137}
  [\href{https://arxiv.org/abs/astro-ph/9712318}{{\ttfamily
  astro-ph/9712318}}].

\bibitem{delos2019b}
M.S.~{Delos}, \emph{{Tidal evolution of dark matter annihilation rates in
  subhalos}}, \href{https://doi.org/10.1103/PhysRevD.100.063505}{\emph{\prd}
  {\bfseries 100} (2019) 063505}
  [\href{https://arxiv.org/abs/1906.10690}{{\ttfamily 1906.10690}}].

\bibitem{planck2018}
{Planck Collaboration}, N.~{Aghanim}, Y.~{Akrami}, M.~{Ashdown}, J.~{Aumont},
  C.~{Baccigalupi} et~al., \emph{{Planck 2018 results. VI. Cosmological
  parameters}}, \href{https://doi.org/10.1051/0004-6361/201833910}{\emph{\aap}
  {\bfseries 641} (2020) A6}
  [\href{https://arxiv.org/abs/1807.06209}{{\ttfamily 1807.06209}}].

\bibitem{despali2016}
G.~{Despali}, C.~{Giocoli}, R.E.~{Angulo}, G.~{Tormen}, R.K.~{Sheth}, G.~{Baso}
  et~al., \emph{{The universality of the virial halo mass function and models
  for non-universality of other halo definitions}},
  \href{https://doi.org/10.1093/mnras/stv2842}{\emph{\mnras} {\bfseries 456}
  (2016) 2486} [\href{https://arxiv.org/abs/1507.05627}{{\ttfamily
  1507.05627}}].

\bibitem{jiang2014}
F.~{Jiang} and F.C.~{van den Bosch}, \emph{{Statistics of Dark Matter
  Substructure: I. Model and Universal Fitting Functions}},
  \href{https://doi.org/10.48550/arXiv.1403.6827}{\emph{arXiv e-prints} (2014)
  arXiv:1403.6827} [\href{https://arxiv.org/abs/1403.6827}{{\ttfamily
  1403.6827}}].

\bibitem{ishiyama2021}
T.~{Ishiyama}, F.~{Prada}, A.A.~{Klypin}, M.~{Sinha}, R.B.~{Metcalf},
  E.~{Jullo} et~al., \emph{{The Uchuu simulations: Data Release 1 and dark
  matter halo concentrations}},
  \href{https://doi.org/10.1093/mnras/stab1755}{\emph{\mnras} {\bfseries 506}
  (2021) 4210} [\href{https://arxiv.org/abs/2007.14720}{{\ttfamily
  2007.14720}}].

\bibitem{limousin2005}
M.~{Limousin}, J.-P.~{Kneib} and P.~{Natarajan}, \emph{{Constraining the mass
  distribution of galaxies using galaxy-galaxy lensing in clusters and in the
  field}},
  \href{https://doi.org/10.1111/j.1365-2966.2004.08449.x}{\emph{\mnras}
  {\bfseries 356} (2005) 309}
  [\href{https://arxiv.org/abs/astro-ph/0405607}{{\ttfamily
  astro-ph/0405607}}].

\bibitem{sereno2016}
M.~{Sereno}, C.~{Fedeli} and L.~{Moscardini}, \emph{{Comparison of weak lensing
  by NFW and Einasto halos and systematic errors}},
  \href{https://doi.org/10.1088/1475-7516/2016/01/042}{\emph{\jcap} {\bfseries
  2016} (2016) 042} [\href{https://arxiv.org/abs/1504.05183}{{\ttfamily
  1504.05183}}].

\bibitem{minor2021}
Q.~{Minor}, M.~{Kaplinghat}, T.H.~{Chan} and E.~{Simon}, \emph{{Inferring the
  concentration of dark matter subhaloes perturbing strongly lensed images}},
  \href{https://doi.org/10.1093/mnras/stab2209}{\emph{\mnras} {\bfseries 507}
  (2021) 1202} [\href{https://arxiv.org/abs/2011.10629}{{\ttfamily
  2011.10629}}].

\bibitem{diamanti2015}
R.~{Diamanti}, M.E.C.~{Catalan} and S.~{Ando}, \emph{{Dark matter protohalos in
  a nine parameter MSSM and implications for direct and indirect detection}},
  \href{https://doi.org/10.1103/PhysRevD.92.065029}{\emph{\prd} {\bfseries 92}
  (2015) 065029} [\href{https://arxiv.org/abs/1506.01529}{{\ttfamily
  1506.01529}}].

\bibitem{dai2018}
B.~{Dai}, B.E.~{Robertson} and P.~{Madau}, \emph{{Around the Way: Testing
  {\ensuremath{\Lambda}}CDM with Milky Way Stellar Stream Constraints}},
  \href{https://doi.org/10.3847/1538-4357/aabb06}{\emph{\apj} {\bfseries 858}
  (2018) 73} [\href{https://arxiv.org/abs/1804.00669}{{\ttfamily 1804.00669}}].

\bibitem{carlberg2020}
R.G.~{Carlberg}, \emph{{The Density Structure of Simulated Stellar Streams}},
  \href{https://doi.org/10.3847/1538-4357/ab61f0}{\emph{\apj} {\bfseries 889}
  (2020) 107}.

\bibitem{bonaca2020}
A.~{Bonaca}, C.~{Conroy}, D.W.~{Hogg}, P.A.~{Cargile}, N.~{Caldwell},
  R.P.~{Naidu} et~al., \emph{{High-resolution Spectroscopy of the GD-1 Stellar
  Stream Localizes the Perturber near the Orbital Plane of Sagittarius}},
  \href{https://doi.org/10.3847/2041-8213/ab800c}{\emph{\apjl} {\bfseries 892}
  (2020) L37} [\href{https://arxiv.org/abs/2001.07215}{{\ttfamily
  2001.07215}}].

\bibitem{yang2020}
S.~{Yang}, X.~{Du}, A.J.~{Benson}, A.R.~{Pullen} and A.H.G.~{Peter}, \emph{{A
  new calibration method of sub-halo orbital evolution for semi-analytic
  models}}, \href{https://doi.org/10.1093/mnras/staa2496}{\emph{\mnras}
  {\bfseries 498} (2020) 3902}
  [\href{https://arxiv.org/abs/2003.10646}{{\ttfamily 2003.10646}}].

\bibitem{jiang2021}
F.~{Jiang}, A.~{Dekel}, J.~{Freundlich}, F.C.~{van den Bosch}, S.B.~{Green},
  P.F.~{Hopkins} et~al., \emph{{SatGen: a semi-analytical satellite galaxy
  generator - I. The model and its application to Local-Group satellite
  statistics}}, \href{https://doi.org/10.1093/mnras/staa4034}{\emph{\mnras}
  {\bfseries 502} (2021) 621}
  [\href{https://arxiv.org/abs/2005.05974}{{\ttfamily 2005.05974}}].

\bibitem{moore2004}
B.~{Moore}, S.~{Kazantzidis}, J.~{Diemand} and J.~{Stadel}, \emph{{The origin
  and tidal evolution of cuspy triaxial haloes}},
  \href{https://doi.org/10.1111/j.1365-2966.2004.08211.x}{\emph{\mnras}
  {\bfseries 354} (2004) 522}
  [\href{https://arxiv.org/abs/astro-ph/0310660}{{\ttfamily
  astro-ph/0310660}}].

\bibitem{drakos2019a}
N.E.~{Drakos}, J.E.~{Taylor}, A.~{Berrouet}, A.S.G.~{Robotham} and C.~{Power},
  \emph{{Major mergers between dark matter haloes - I. Predictions for size,
  shape, and spin}}, \href{https://doi.org/10.1093/mnras/stz1306}{\emph{\mnras}
  {\bfseries 487} (2019) 993}
  [\href{https://arxiv.org/abs/1811.12839}{{\ttfamily 1811.12839}}].

\bibitem{penarrubia2012}
J.~{Pe{\~n}arrubia}, A.~{Pontzen}, M.G.~{Walker} and S.E.~{Koposov}, \emph{{The
  Coupling between the Core/Cusp and Missing Satellite Problems}},
  \href{https://doi.org/10.1088/2041-8205/759/2/L42}{\emph{\apjl} {\bfseries
  759} (2012) L42} [\href{https://arxiv.org/abs/1207.2772}{{\ttfamily
  1207.2772}}].

\bibitem{diCintio2014}
A.~{Di Cintio}, C.B.~{Brook}, A.V.~{Macci{\`o}}, G.S.~{Stinson}, A.~{Knebe},
  A.A.~{Dutton} et~al., \emph{{The dependence of dark matter profiles on the
  stellar-to-halo mass ratio: a prediction for cusps versus cores}},
  \href{https://doi.org/10.1093/mnras/stt1891}{\emph{\mnras} {\bfseries 437}
  (2014) 415} [\href{https://arxiv.org/abs/1306.0898}{{\ttfamily 1306.0898}}].

\bibitem{read2019}
J.I.~{Read}, M.G.~{Walker} and P.~{Steger}, \emph{{Dark matter heats up in
  dwarf galaxies}}, \href{https://doi.org/10.1093/mnras/sty3404}{\emph{\mnras}
  {\bfseries 484} (2019) 1401}
  [\href{https://arxiv.org/abs/1808.06634}{{\ttfamily 1808.06634}}].

\bibitem{deLeo2023}
M.~{De Leo}, J.I.~{Read}, N.E.D.~{No{\"e}l}, D.~{Erkal}, P.~{Massana} and
  R.~{Carrera}, \emph{{Surviving the waves: evidence for a dark matter cusp in
  the tidally disrupting Small Magellanic Cloud}},
  \href{https://doi.org/10.1093/mnras/stae2428}{\emph{\mnras} {\bfseries 535}
  (2024) 1015} [\href{https://arxiv.org/abs/2303.08838}{{\ttfamily
  2303.08838}}].

\bibitem{matthias2006}
M.~{Hoeft}, G.~{Yepes}, S.~{Gottl{\"o}ber} and V.~{Springel}, \emph{{Dwarf
  galaxies in voids: suppressing star formation with photoheating}},
  \href{https://doi.org/10.1111/j.1365-2966.2006.10678.x}{\emph{\mnras}
  {\bfseries 371} (2006) 401}
  [\href{https://arxiv.org/abs/astro-ph/0501304}{{\ttfamily
  astro-ph/0501304}}].

\bibitem{noh2014}
Y.~{Noh} and M.~{McQuinn}, \emph{{A physical understanding of how reionization
  suppresses accretion on to dwarf haloes}},
  \href{https://doi.org/10.1093/mnras/stu1412}{\emph{\mnras} {\bfseries 444}
  (2014) 503} [\href{https://arxiv.org/abs/1401.0737}{{\ttfamily 1401.0737}}].

\bibitem{numpy}
C.R.~Harris, K.J.~Millman, S.J.~van~der Walt, R.~Gommers, P.~Virtanen,
  D.~Cournapeau et~al., \emph{Array programming with {NumPy}},
  \href{https://doi.org/10.1038/s41586-020-2649-2}{\emph{Nature} {\bfseries
  585} (2020) 357}.

\bibitem{matplotlib}
J.D.~Hunter, \emph{Matplotlib: A 2d graphics environment},
  \href{https://doi.org/10.1109/MCSE.2007.55}{\emph{Computing in Science \&
  Engineering} {\bfseries 9} (2007) 90}.

\bibitem{scipy}
P.~Virtanen, R.~Gommers, T.E.~Oliphant, M.~Haberland, T.~Reddy, D.~Cournapeau
  et~al., \emph{{{SciPy} 1.0: Fundamental Algorithms for Scientific Computing
  in Python}}, \href{https://doi.org/10.1038/s41592-019-0686-2}{\emph{Nature
  Methods} {\bfseries 17} (2020) 261}.

\bibitem{colossus}
B.~{Diemer}, \emph{{COLOSSUS: A Python Toolkit for Cosmology, Large-scale
  Structure, and Dark Matter Halos}},
  \href{https://doi.org/10.3847/1538-4365/aaee8c}{\emph{\apjs} {\bfseries 239}
  (2018) 35} [\href{https://arxiv.org/abs/1712.04512}{{\ttfamily 1712.04512}}].

\end{thebibliography}\endgroup

\end{document}